# The phase diagrams of iron-based superconductors: theory and experiments

## Les diagrammes de phases des supraconducteurs à base de fer: La théorie etles expériences

A. Martinelli[1,*], F. Bernardini[2,3], S. Massidda[3]

[1] SPIN-CNR, C.so Perrone 24, I-16152 Genova – Italy

[2] Dipartimento di Fisica, Università di Cagliari, Cittadella Universitaria, I-09042 Monserrato - Italy

[3] IOM-CNR c/o Dipartimento di Fisica, Università di Cagliari, Cittadella Universitaria, I-09042 Monserrato - Italy

## Abstract

Phase diagrams play a primary role in the understanding of materials properties. For iron-based superconductors (Fe-SC), the correct definition of their phase diagrams is crucial because of the close interplay between their crystallo-chemical and magnetic properties, on one side, and the possible coexistence of magnetism and superconductivity, on the other.

The two most difficult issues for understanding the Fe-SC phase diagrams are: 1) the origin of the structural transformation taking place during cooling and its relationship with magnetism; 2) the correct description of the region where a crossover between the magnetic and superconducting electronic ground states takes place. Hence a proper and accurate definition of the structural, magnetic and electronic phase boundaries provides an extremely powerful tool for material scientists.

For this reason, an exact definition of the thermodynamic phase fields characterizing the different structural and physical properties involved is needed, although it is not easy to obtain in many cases. Moreover, physical properties can often be strongly dependent on the occurrence of micro-structural and other local-scale features (lattice micro-strain, chemical fluctuations, domain walls, grain boundaries, defects), which, as a rule, are not described in a structural phase diagram.

In this review, we critically summarize the results for the most studied 11-, 122- and 1111-type compound systems, providing a correlation between experimental evidence and theory.

## Résumé

Les diagrammes de phase jouent un rôle de première importance dans la compréhension des propriétés des matériaux. En ce qui concerne les supraconducteurs à base de fer (Fe-SC), la définition correcte de leurs diagrammes de phase est cruciale à cause de l'intime interaction entre leurs propriétés cristallochimiques et magnétiques, d'un côté, et la possible coexistence de magnétisme et de supraconductivité, de l'autre.

---

[*] Correspong author: alberto.martinelli@spin.cnr.it

Les deux difficultés principales pour la compréhension des diagrammes de phase Fe-SC sont: 1) l'origine de la transformation structurelle ayant lieu pendant le refroidissement et sa relation avec le magnétisme; 2) la description correcte de la région où survient un recouvrement entre les états fondamentaux électroniques, magnétique et supraconducteur électronique survient. De ce fait, une définition appropriée et précise des frontières des phases structurelle, magnétique et électronique fournit un outil extrêmement puissant pour les scientifiques du domaine des matériaux.

Pour cette raison, une définition exacte des champs de phases thermodynamiques caractérisant les différentes propriétés structurelles et physiques impliquées est nécessaire, bien qu'elle ne soit pas aisée à obtenir dans de nombreux cas. De plus, les propriétés physiques peuvent souvent dépendre fortement de la survenue de caractéristiques micro-structurelles ou autres à l'échelle locale (micro-contraintes dans le réseau, fluctuations chimiques, parois de domaines, joints de grains, défauts), qui, d'ordinaire, ne sont pas décrites dans un diagramme de phases structurelles.

Dans cette revue, nous résumons de manière critique les résultats obtenus pour les systèmes composites les plus étudiés de types 11-, 122-and 1111-type, qui établissent une corrélation entre les preuves expérimentales et la théorie.

KEYWORDS: iron-based superconductors; phase diagrams; structural transformations; superconductivity; magnetism; nematicity

MOTS-CLÉS: supraconducteurs à base de fer; diagrammes de phase; transformations structurels; supraconductivité; magnétisme; phase nématique

## 1. Overview

The fascinating physics distinguishing the class of materials referred to as iron-based superconductors (Fe-SC) emerges from the delicate and tangled interplay between magnetism, superconductivity and crystallo-chemistry. The understanding of the normal state properties is a fundamental step in the development of a theory of superconductivity in Fe-SC. Despite the outstanding attention paid to these systems, it is not yet clear if a universal phase diagram can be established.

Pnictides (typically 122- and 1111-type systems) display quite similar structural and magnetic phase relationships. The chalcogenides (11-type systems) are rather different; in particular, the pseudo-binary system $\beta$-FeSe$_{1-x}$ - $\beta$-Fe$_{1+y}$Te is the only relevant system among the Fe-chalcogenides and will be treated in detail below. A schematic phase diagram is drawn in Figure 1.1, highlighting the features shared by most of these systems.

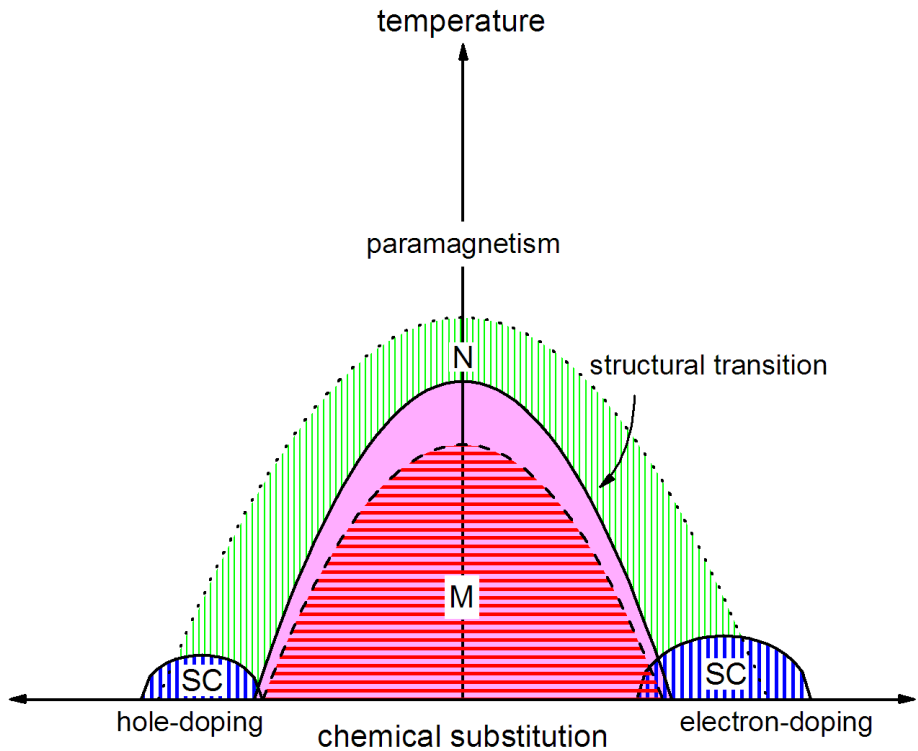

Figure 1.1: Schematic phase diagram of Fe-based superconductors, showing the most credited phase relationships, with nematic (N), magnetic (M) and superconducting (SC) phases labelled; in the case of isovalent substitution no doping occurs.

## 1.1 Structural transformations

As a rule, the undoped parent compound undergoes a structural transformation upon cooling at the temperature $T_s$, followed by a magnetic transition at $T_m$, where $T_m \lesssim T_s$. For the 122- and 1111-type compounds, a translation-equivalent (*translationengleiche*) structural transition of index 2 changes the structure from tetragonal to orthorhombic. The unit cell of the low-temperature orthorhombic phase is rotated by 45° in the $xy$ plane with respect to that of the high-temperature tetragonal one, and the edges of the basal cell are a factor of $\overline{\phantom{x}}$ larger in the orthorhombic structure.

So far, two main scenarios have been proposed to explain the occurrence of the transformation from tetragonal to orthorhombic in the 122- and 1111-type compounds: 1) orbital ordering drives the structural transition and induces magnetic anisotropy, thus triggering the magnetic transition [1,2,3]; 2) magnetic fluctuations drive the structural transition and induce orbital ordering [4].

The structural transformation temperature $T_s$ can be reliably ascertained by diffractometric analysis carried out as a function of temperature; in particular for 122- and 1111-type compounds a selective Bragg peak splitting marks the symmetry breaking (Figure 1.2). Conversely, no anomaly can be detected on crossing $T_s$ by optical measurements [5,6], since no displacive optical mode is involved [7]. For this reason in this review we usually refer to the structural transformations temperatures $T_s$ obtained by diffraction, whenever not specified. Otherwise it is indicated when data stems from other trustworthy methods, such as specific heat measurements or NMR analysis.

In this review these kinds of data are used to draw phase boundaries in phase diagrams. Remarkably the thermal dependence of the resistivity often exhibits discontinuities that are commonly related to the structural transition; actually such discontinuities mark a change of the electronic properties, rather than a real structural change, that in some cases can become extremely reduced. Hence these kinds of data are not considered in this review, since they cannot be considered a reliable probe for detecting structural transformations.

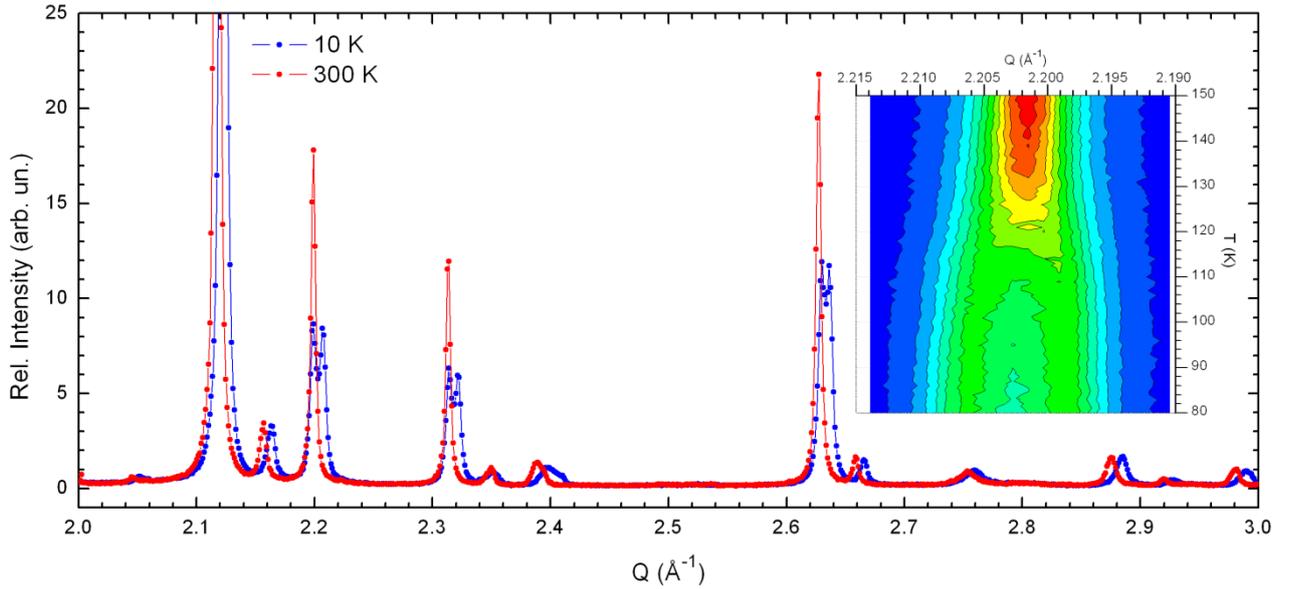

Figure 1.2: Superposition of the powder diffraction patterns of La(Fe$_{0.90}$Ru$_{0.10}$)AsO collected at 300 K and 10 K showing the selective peak splitting affecting the Bragg peaks with strong components in the *ab* plane on cooling marking the tetragonal to orthorhombic structural transformation (X-ray synchrotron data); the inset shows the thermal evolution of the tetragonal 110 diffraction line splitting on cooling into the orthorhombic 200 and 020 lines.

### 1.2 Magnetism

When dealing with magnetic ordering in 1111- and 122-type compounds, confusion can arise when comparing works referring to the parent tetragonal phase with those analyzing the distorted orthorhombic structure. In fact, due to the aforementioned rotation undergone by the unit cell after the structural transformation, the in-plane magnetic wave-vector is (1,0) or (½,½) when referred to the orthorhombic or the tetragonal structure (Figure 1.3), respectively (for a detailed discussion, see ref. [8]). Figure 1.4 shows the typical spin orderings characterizing the prototypical Fe$_{1+y}$Te, BaFe$_2$As$_2$ and LaFeAsO compositions in their low-temperature polymorphic modifications.

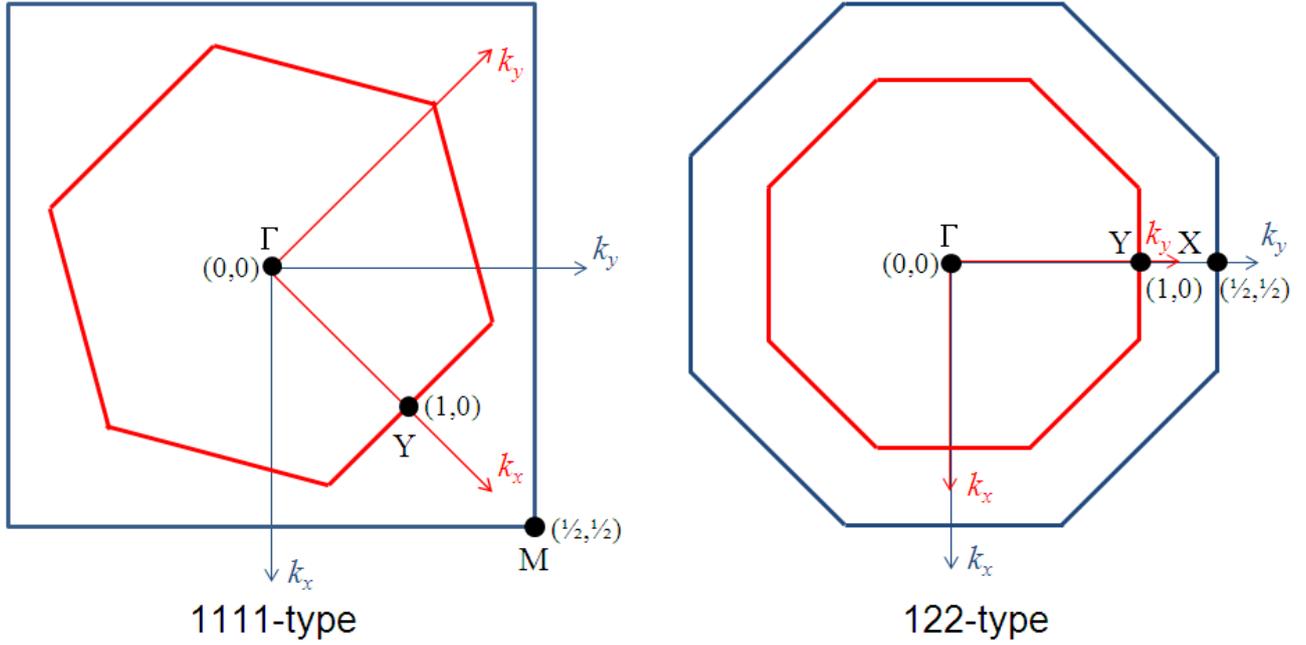

Figure 1.3: Superpositions of the Brillouin zones in the $k_x$- $k_y$ plane of the Fe layers for the *P4/nmm* and *Cmme* space groups pertaining to the 1111-type compounds (on the left) and for the *I4/mmm* and *Fmmm* space groups pertaining to the 122-type compounds (on the right).

With a few exceptions, magnetic ordering is always observed in conjunction with the breaking of tetragonal symmetry, but the opposite is not true; conversely, in several cases, symmetry breaking has been observed in fully superconductive compounds where magnetism is completely suppressed. Moreover, the relationship between the thermal dependence of the crystal structure and magnetism appears to be different in the 11- type systems as compared to the 122- and 1111-type systems.

The nature of the ordered magnetic state in these compounds is still debated, and different interpretations were proposed, in particular for 122- and 1111-type compounds [9] (for a recent review on magnetic interactions in Fe-SC, the reader is referred to Bascones *et al.* [10], appearing in this same volume). A first scenario supports an itinerant character of the antiferromagnetic state. This hypothesis is supported by experimental evidence that the nesting wave-vector for the electron and hole Fermi surface pockets $\mathbf{k}_{nesting} = (\pi,\pi)$ is consistent with the in-plane tetragonal magnetic wave-vector $\mathbf{k}_{magnetic} = (\pi,\pi)$ observed in 122- and 1111-type compounds (Figure 1.4). This leads to a spin density wave instability [5,11,12] where the magnitude of the magnetic moment exhibits a sinusoidal modulation with distance. In this context, it is worth noting that the periodicities of a spin density wave are generally not rational fractions of the periodicities of the hosting lattice, however in 122- and 1111-type compounds, spin ordering is always commensurate. Moreover, there is mounting evidence that the superconductivity in these materials is strongly correlated with the in-plane $(\pi,\pi)$ spin fluctuations (for a recent review on spin fluctuations in Fe-SC, the reader is referred to Inosov [13], appearing in this same volume).. Moderate electronic correlations, small magnetic moments in the ordered state and a significant broadening of the spin-wave dispersion at high energies further support this scenario.

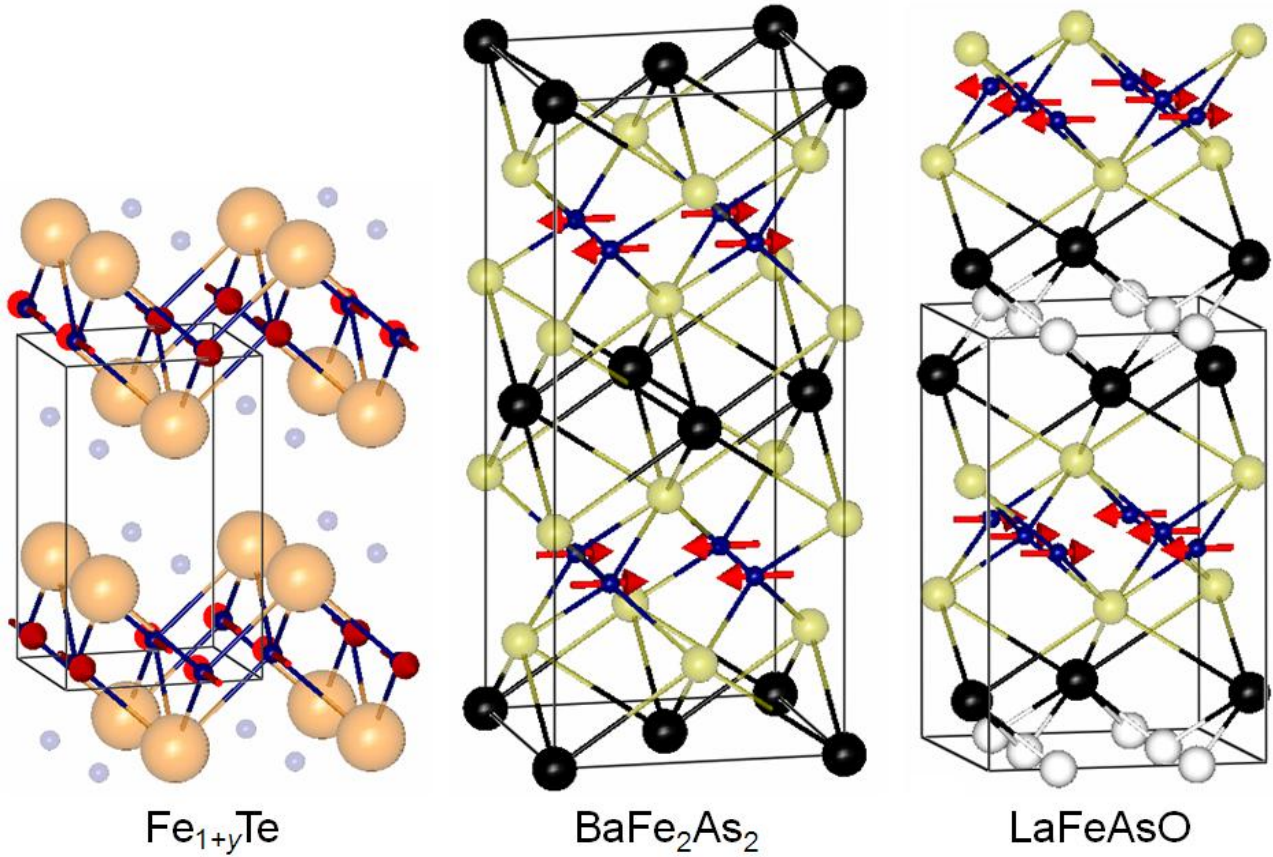

Figure 1.4: Crystal structures and spin orderings characterizing the three most important Fe-SC families.

Alternative theories based on local moments have been proposed, in which magnetic frustration is thought to be induced by near-neighbour and next-to-near-neighbour interactions among local Fe moments [14,15,16], whereas ferro-orbital ordering has been suggested to drive both structural and magnetic transitions [2,17]. A tetragonal-to-orthorhombic distortion is not a surprising behaviour in a magnetic phase with an antiferromagnetic stripe order. Indeed, a square lattice is distorted into a rectangular one where the spins are parallel along the shorter edge and anti-parallel along the longer one. In any case, the experimental evidence shows that $T_s$ is independent on the presence of magnetic fields; this suggests that a common driving force that is independent of the spin degrees of freedom is at the origin of both magnetic and structural transitions. A microscopic theory of the coupled structural and magnetic transitions suggests that the uneven occupation of the $d_{xz}$ and $d_{yz}$ orbitals leads to orbital ordering [2]. At $T > T_s$, both orbitals have the same average occupation and no ordering occurs: a square lattice is energetically stable. At lower temperatures, the higher filling of one of the two orbitals lifts their degeneracy and makes the electron distribution around each Fe atom anisotropic; because of the inter-site Coulomb repulsion, a rectangular lattice becomes preferred.

This scenario foreseeing orbital ordering is in agreement with the results of angle-resolved photoemission experiments using a linear-polarized laser beam, showing that the Fermi surface at the Brillouin-zone center is dominated by a single $d_{xz}$ or $d_{yz}$ orbital at low temperatures [18]. Moreover, this model is able to predict both the nematic behaviour of the resistivity by a Kondo-like scattering behaviour and the effect of doping which, by adding or subtracting charge to the orbitals, makes the charge distribution around the Fe atoms more isotropic and the orbital ordering less favourable. On the other hand, in DFT calculations, the orthorhombic structure results more stable than the tetragonal one only when magnetic order is taken into account

In the $Fe_{1+y}(Te_{1-x}Se_x)$ system, magnetism appears different. Indeed, the nesting wave-vector for the electron and hole Fermi surface pockets is also $\mathbf{k}_{nesting} = (\pi,\pi)$ in this case, but the experimental in-plane magnetic propagation vector is $\mathbf{k}_{magnetic} = (\pi,0)$ (Figures 1.3 and 1.4), suggesting that antiferromagnetism in $Fe_{1+y}Te$ does not originate from the Fermi surface nesting of itinerant charges, but rather from local magnetic moments.

*1.3 Superconductivity*

The highest $T_c$ is achieved when magnetism is completely destroyed or at least strongly hindered; this can usually be obtained by electron- or hole-doping, but in some cases superconductivity can also emerge by applying pressure. At room pressure, superconductivity can be induced (or enhanced) by three different kinds of doping:

1) electron doping: an increase in the negative charge in the FeAs layers is obtained for example in the $Ba(Fe_{1-x}Co_x)_2As_2$, $RE$FeAs($O_{1-x}F_x$) and $RE(Fe_{1-x}Co_x)$AsO systems.

2) hole doping: the positive charge in the FeAs layer is increased in the $(Ba_{1-x}K_x)Fe_2As_2$, $(La_{1-x}Sr_x)$FeAsO and $RE$FeAs($O_{1-x}\square_x$) systems

3) isovalent doping: by replacing Se with Te in $\beta$-$FeSe_{1-x}$, Fe with Ru in the 122 systems such as $Ba(Fe_{1-x}Ru_x)_2As_2$, As with P in $BaFe_2(As_{1-x}P_x)_2$ and $LaFe(As_{1-x}P_x)O$ systems.

The destabilization of the magnetic ground state thus appears as a key factor for favouring superconductivity in these materials, although, as mentioned above, the prevailing scenario suggests that the electron-pairing interactions are produced by the same magnetic interactions that are driving magnetic ordering. In particular, superconductivity arises after the complete suppression of magnetic ordering in some systems (left side of the phase diagram, Figure 1.1), whereas in other cases, the coexistence between these two ground states is seen within an under-doped region (right side of the phase diagram, Figure 1.1). Hence, the correct characterization of the region where a crossover between magnetism and superconductivity takes place is fundamental (whether there is phase separation or phase coexistence).

Another key factor for obtaining superconductivity seems to be the suppression of the structural transformation observed during cooling. In some systems, superconductivity can only be observed when the tetragonal symmetry at room temperature is retained during cooling by appropriate substitution/doping (left side of the phase diagram in Fig. 1.1). In other cases, a fully superconductive state can also be obtained after reduction of symmetry, even though the highest $T_c$ is generally obtained after the complete suppression of symmetry breaking (right side of the phase diagram in Fig. 1.1).

*1.4 Nematicity*

Electronic and magnetic nematic degrees of freedom are believed to be associated with the origin of structural transitions and their interplay with magnetism in 122- and 1111-type systems, and with the unconventional superconductivity in Fe-based materials. Nematic order consists in the spontaneous breaking of the electronic symmetry between the $x$ and $y$ directions in the Fe-plane, but not of the underlying (tetragonal) lattice. As a consequence, in the nematic state, several physical properties (transport, magnetic and optical properties) display a symmetry that is different from the one seen in the crystal lattice. Nematic fluctuations thus develop during cooling at $T_n > T_s$ in under-doped compounds, or even in optimally-doped compounds showing no evidence of symmetry breaking (Figure 1.1). Unfortunately, in the literature of Fe-based superconductors, the terms "nematic (phase?)" and "orthorhombic phase" are often used as synonyms. In this context, a clarification is necessary: the space group (and consequently the pertaining crystal system) of any crystalline phase is not defined by the metric of its lattice constants, but rather by its lattice symmetry. For example, the tetragonal symmetry does impose equality of the linear parameters $a$ and $b$; in the orthorhombic system, $a$ can be different than $b$, but in principle $a = b$ is still consistent when symmetry properties along the $a$ and $b$ axes differ. In the Fe-SC literature, the nematic phase is usually said to occur as the rotational symmetry in the Fe-plane is broken ($C_4 \rightarrow C_2$ point-group reduction), but not the translational one (the crystal lattice maintains a tetragonal-like metric). In fact, the nematic phase is orthorhombic from the symmetry point of view, since the 4-fold rotational axis is suppressed, even though the lattice maintains a tetragonal lattice metric. In particular, by suppressing the generator 4-fold rotational axis in the tetragonal *P*4/*nmm* space group, the orthorhombic *Cmme* space group is obtained [19]. Nonetheless, it is commonly assumed that a nematic phase occurs when the rotational and translational symmetries are broken at separate transitions. Nematic fluctuations should therefore occur within an underlying pseudo-tetragonal lattice, in which no orthorhombic crystallographic distortion can be seen. The nematic state most likely has an electronic origin and is driven by the same fluctuations that induce superconductivity and magnetic ordering [20]. Unfortunately, the field lines defining the nematic phase are known for very few systems, since in most studies, experimental data on the nematic phase are only available for a few compositions.

## 1.5 Relationship between magnetism and superconductivity

One of the most prominent issues in the physics of Fe-SC is the interplay between the magnetic and superconducting order parameters when charge doping, pressure or other parameters are modified. In this context, it is imperative to precisely identify the intrinsic microscopic properties in the cross-over region between the magnetic and superconductive ground states. In this paper, we will use the terms "(phase) coexistence" to indicate that the magnetic and superconductive order parameters are finely intertwined at the nanoscopic level, possibly coexisting in the same nanoscopic volume [21]. Conversely, the term "(phase) segregation" indicates that the volumes of the magnetic and superconducting phases are demixed at the meso- to macroscopic levels.

We note that the relationship between the magnetic and superconducting order parameters can be strongly dependent on the quality of the analyzed sample; hence samples characterized by the same nominal composition can actually exhibit segregation or coexistence, depending on their quality grade.

## 1.6 Quantum critical point

Another fascintaing features in Fe-SC diagrams is the quantum phase transition at the boundary between the antiferromagnetic and superconductive states, where one can control the competition between these two ground states by tuning selected parameters, such as magnetic field, pressure, electron density and composition.

We still do not know whether a quantum critical phase lies beneath the superconducting dome, or whether the criticality is avoided by the transition to the superconducting state [22]. Quantum criticality appears when the critical temperature of a phase transition goes to zero. This is indeed the case for doped Fe-SC where sufficient doping (holes or electrons) drives $T_m$ to 0 K. The presence of this quantum critical point (QCP) is suggested by the anomalous behaviour in the resistivity (linear instead of quadratic behaviour) versus temperature, and by the breaking of Kohler's rule in the magneto-resistance, in the region of the phase diagram just above the end points of the antiferromagnetic phase dome. Quantum criticality is not a feature unique to Fe-SC, as it is found in heavy-fermions and superconducting cuprates (Cu-SC). What makes the quantum criticality important in Fe-SC is the possible strong overlap between superconductivity and the quantum critical phase. Indeed, in Fe-SC, the position of the magnetic QCP, as extrapolated from the shape of the antiferromagnetic dome, is near the maximum $T_c$. This implies that a strong overlap, not present in the Cu-SC, may exist between the quantum critical region and the superconducting phase. The origin of the quantum criticality is not clear; a major obstacle to probing the presence or absence of a QCP inside the superconducting dome is the presence of superconductivity itself, which makes most experimental probes insensitive to its presence. Unfortunately, removing superconductivity by magnetic fields would affect the magnetic phase diagram. The study of the coexistence of quantum criticality and superconductivity needs to resort to different techniques. In this respect, iso-electronic substitution, such as P-substitution, can be used to study the quantum critical behaviour in Fe-SC.

*1.7 Final remarks*

The characterization of the crystallographic phase boundaries constituting the phase diagram is of critical importance, especially when a tetragonal-to-orthorhombic transformation occurs. In this context, we must point out that the *a*- and *b*-axes of the orthorhombic phase differ by less than 1% in pure 122- and 1111-type compounds. In addition, in most cases, chemical substitution progressively reduces the orthorhombic distortion of the structure, such as in the hole-doped 122- and electron-doped 1111-type systems. Therefore, the detection of the structural transformation and the exact determination of $T_s$ can be strongly affected by the instrumental resolution of the diffractometer, as well as by the accuracy of the structural analysis, which can lead to contradictory definitions of the critical temperatures for a given composition. It is also not unusual for different research groups to find different values for the transition temperature $T_s$, for a given nominal composition. Indeed, most of the studied materials are solid solutions and samples that are prepared in different laboratories and can be affected by notable compositional deviations (and possibly compositional fluctuations), despite being characterized by the same nominal composition. We point out that the $T_s$, $T_m$, and $T_c$ values we refer to in the text are obtainedfrom reliable experimental techniques that directly probe the investigated property; in many cases we explicitly state how these values are obtained. Otherwise, for the sake of clarity, we refer the reader to the bibliography in which a more complete description of the applied experimental techniques can be found.

Finally, it is of critical importance to verify whether the phase diagram complies with the phase rule and other thermodynamic principles that control the relationships among the different phases [23,24,25]. In particular, boundaries that seem very unlikely should be carefully examined, and the nature of the structural transitions should be verified whenever possible, in order to properly separate the different stability phase fields. Unfortunately data reported in literature are often incomplete from this point of view; for this reason in many cases it is not possible to draw equilibrium curves defining the 2-phase field that must be present when a 1[st] order transition takes place.

Here we focus on the phase diagrams of the 11-, 122- and 1111-type compounds, since they have been studied in more depth, and critically evaluate the experimental results reported in the literature; in some cases, the phase diagrams will be tentatively re-drawn after re-assessing some of the published experimental data.

## 2. The 11-type systems

We provide a selective overview of the properties of the $Fe_{1+y}(Te_{1-x}Se_x)$ system, which has been extensively studied. On the other hand, the $Fe_{1+y}(Te_{1-x}S_x)$ and $Fe(Se_{1-x}S_x)$ systems did not stimulate much interest, and for this reason will not be treated.

### 2.1 The $Fe_{1+y}(Te_{1-x}Se_x)$ system

Fe-chalcogenide superconductors are of great interest because they are the simplest Fe-based superconductors. Furthermore, the superconducting properties of Fe-chalcogenides are strongly affected by pressure. At optimal doping ($x \sim 0.5$), $Fe_{1+y}(Te_{1-x}Se_x)$ has, until now, shown to have the highest superconducting critical temperature ($T_c = 15.6$ K) among chalcogenides at zero pressure. $\beta$-FeSe shows a $T_c$ that strongly depends on external pressure: $T_c$ increases from 8 K at ambient pressure up to 37 K at $p \sim 9$ GPa; on the other hand, $Fe_{1+y}Te$ is a noticeable example of a non-superconducting parent compound.

The two end-members of the $Fe_{1+y}(Te_{1-x}Se_x)$ system are characterized by extremely similar structures. Both crystallize in the $P4/nmm$ -129 space group at room temperature, but they are not isotypic; in addition, the thermal dependence of their structures also displays significant differences.

### 2.1.1 The $\beta$-$Fe_{1+y}Te$ end member

At room temperature, the $\beta$-$Fe_{1+y}Te$ end-member crystallizes into a strongly defective $Cu_2Sb$ structure-type, where Fe atoms are located at two different structural sites: the tetrahedral $2a$ and the interstitial $2c$ Wyckoff sites. The occupancy range of the interstitial site has not yet been defined: at present, it is reported to extend up to $y = 0.30$ [26], but further studies are needed. During cooling, different transitions take place, depending on the amount of Fe at the interstitial site quantified by the parameter $y$ [26,27,28,29]. For $y \leq 0.10$, a 1[st] order $P4/nmm \rightarrow P2_1/m$ structural transition takes place around 70 K, coupled with a simultaneous 1[st] order magnetic transition; neutron diffraction analyses reveal that the resulting antiferromagnetic structure is characterized by a magnetic propagation vector $\mathbf{k} = (\frac{1}{2},0,\frac{1}{2})$ with an ordered Fe moment of $\sim 1.9$ - $2.5$ $\mu_B$, which decreases with the increase in the interstitial Fe concentration [27,30,31,32]. Fe-moments are oriented along the shorter $b$-axis in the $a$-$b$ plane and form an ordered double-stripe structure (Figure 1.4). In addition, the existence of this double-stripe antiferromagnetic structure is confirmed by theoretical DFT calculations, which show that this structure results from the interactions between near-, next-near- and next-next-nearest-neighbors [33,34,35,36,37]. At the tricritical point ($y = 0.11$), the eutectoid transformation $P4/nmm \rightarrow P2_1/m + Pmmn$ takes place, since the compositions of all three phases must be different for an invariant reaction under thermodynamic equilibrium. For $0.11 < y < 0.14$, β-Fe$_{1+y}$Te undergoes two structural transitions, a higher-temperature 2$^{nd}$ order $P4/nmm \rightarrow Pmmn$ transformation, followed by the 1$^{st}$ order $Pmmn \rightarrow P2_1/m$ one at lower temperature. We note that the $Pmmn \rightarrow P2_1/m$ transformation is not complete, but that the orthorhombic and monoclinic polymorphs coexist at low temperature [28,29,26,38]. For $y \geq 0.14$, a 2$^{nd}$ order $P4/nmm \rightarrow Pmmn$ transition occurs around 60 K, coupled with a short-range incommensurate magnetic ordering characterized by a temperature-dependent incommensurate propagation vector $\mathbf{k} = (\pm\delta,0,\frac{1}{2})$. Hence in the magnetic structure, the δ-value can be tuned by varying the amount of interstitial Fe up to $\delta = 0.5$, where a commensurate ordering sets in [27,32]. The sequence of structural transformations occurring in β-Fe$_{1+y}$Te can be understood by symmetry mode analysis: the $P4/nmm$ $\rightarrow Pmmn$ transformation is obtained by the condensation of the distorsive $A_{2g}$ soft mode, whereas for the $P4/nmm \rightarrow P2_1/m$ transition, the distorsive $E_g$ soft mode must also be active. The $E_g$ mode is hindered by the increase in the Fe content, down to its complete suppression, whereas the $A_{2g}$ mode is much less affected by stoichiometry [7]. For this reason, above a critical Fe content, a two-step structural phase transition is observed during cooling for $0.11 < y < 0.14$, where the $A_{2g}$ mode condensates at a higher temperature than the $E_g$ mode. For higher Fe content, the $E_g$ mode is completely suppressed and only the $A_{2g}$ mode is active, thus leading to the formation of an orthorhombic structure. Figure 2.1 shows a tentative phase diagram at normal pressure, drawn for the β-Fe$_{1+y}$Te system by assessing the data reported in ref. [28,29,31].

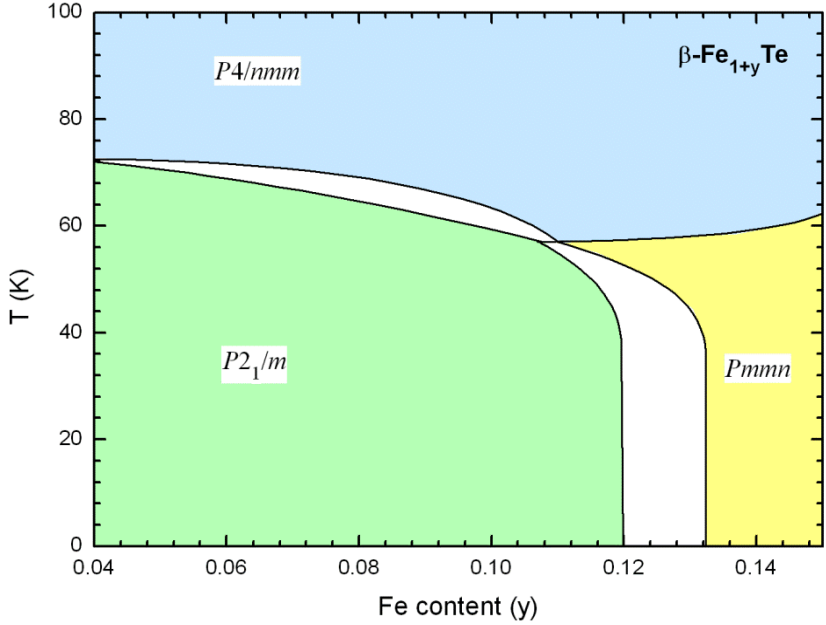

Figure 2.1: Assessed $\beta$-Fe$_{1+y}$Te phase diagram (redrawn from data reported in ref. [28,29,31]).

A few studies were carried out under pressure, with the aim of understanding whether pressure can suppress magnetic order and enhance spin fluctuations, eventually leading to superconductivity: in none of the cases did superconductivity result from applying an external pressure. In Fe$_{1.05}$Te, at room temperature, a pressure of 4 GPa induces a lattice collapse that is not symmetry-breaking [39]. A subsequent study of a sample with the Fe$_{1.087}$Te composition revealed the existence of a similar collapsed tetragonal phase in the $4.1 \leq p \leq 10$ GPa pressure range, whereas a more compressible tetragonal phase was detected for $10 \leq p \leq 16.6$ GPa; further compression led to an amorphous phase [40]. For the same composition, the monoclinic structure was found to be stable at low temperatures and at pressures up to $\sim 1.2$ GPa, whereas an orthorhombic phase (*Pmmn* space group) was observed between 50 K and 60 K at $\sim 1.2$ GPa, and is characterized by an incommensurate antiferromagnetic ordering; above 60 K and up to $\sim 1.2$ GPa, no structural change affects the tetragonal phase [41]. In a sample with the Fe$_{1.03}$Te composition, the commensurate antiferromagnetic order weakened with increasing pressure up to $\sim 2$ GPa, when the system became a low-temperature bulk ferromagnet [42]. A sample with the Fe$_{1.08}$Te composition showed different structural transformations during cooling under different applied pressures [43]: a *P4/nmm* → *P2$_1$/m* transformation was found to occur around $\sim 55$ K at $\sim 0.5$ GPa, while at $\sim 1.4$ GPa, it is replaced by a *P4/nmm* → *Pmmn* transition. At higher pressures, a *P4/nmm* → *P4/nmm* structural transformation is observed, with a transition temperature that increases with pressure, from $\sim 6.0$ K at 2.29 GPa to $\sim 90$ K at 2.9 GPa.

Theoretical studies were carried out using the density functional theory approach [44]; those calculations do not consider deviations from perfect stoichiometry and are not very accurate in the determination of the pressure values of the phase transitions. Nevertheless, they shed light on the behaviour of magnetic order versus pressure. Calculations show that, starting from the monoclinic double-stripe antiferromagnetic order, a first transition leads to a tetragonal ferromagnetic structure at 2.1 GPa, in good agreement with some experimental results [42,43]. At higher pressures, $\beta$-$Fe_{1+y}Te$ undergoes a series of phase transitions between the NiAs and MnP structures with a ferromagnetic ground state (which was never actually observed experimentally), until the magnetization goes to zero as the pressure increases above $p = 17$ GPa. It is interesting to note that the magnetic order that is in place just before the quenching of magnetism is ferromagnetic. The $\beta$-$Fe_{1+y}Te$ compound is speculated to not show any superconducting behaviour under pressure due to the absence of an AFM order, which, under the effect of pressure, transforms into an AFM spin fluctuation.

### 2.1.2 The $\beta$-FeSe$_{1-x}$ end member

The end-member $\beta$-$FeSe_{1-x}$ is isotypic with $\alpha$-PbO [45]. In some recent publications, this phase is improperly referred to as $\alpha$-$FeSe_{1-x}$, but in the Fe - Se system, the $\alpha$-phase is actually a polymorph of $Fe_7Se_8$ [45]. $\beta$-$FeSe_{1-x}$ is not stoichiometric and exhibits a slight defective Se sub-lattice [45,46,47,48]; careful analyses showed that this phase has a very narrow compositional range of $0.963 \leq x \leq 0.994$ [47]. Puzzling results were obtained for samples containing slightly larger amounts of Se, where a slow conversion of $\beta$-$FeSe_{1-x}$ into the NiAs-type $\delta$-$FeSe_{1-x}$ phase ($P6_3/mmc$ space group [45]) is reported to occur below $\sim 573$ K [49]. In the binary Fe-Se phase diagram, the $\delta$-phase actually decomposes below 623 K by the eutectoid reaction $\delta \rightarrow \beta + \gamma$', where $\gamma$' is a polymorph of $Fe_7Se_8$, whose structure is a monoclinic deformation of the NiAs-type structure [45]. The examined composition likely contained an exceeding amount of Se, leading to a thermodynamic instability during the thermal treatment and the formation of the $\gamma$' phase.

During cooling, the $\beta$-$FeSe_{1-x}$ phase undergoes a $P4/nmm \rightarrow Cmme$ structural transition, characterized by a continuous variation of the cell volume [46,47]; rather different $T_s$ are reported, ranging from 100 K down to 70 K [46,47,50,51]. No hysteretic behaviour is observed, supporting the conclusion that the transition is $2^{nd}$ order [47], although deeper analysis would be needed. Recent analyses ascertained that magnetic fluctuations cannot drive the structural transition, since they set in only below $T_s$ [52], but symmetry breaking is rather originated by orbital degrees of freedom [53]. Remarkably, ARPES measurements detected a degeneracy removal of the $d_{xz}$ and $d_{yz}$ orbitals at the $\Gamma$/Z and M points; in particular the splitting at the M point was found to be closely related to the structural transition, showing the existence of a $d$-wave orbital ordering [54].

The orthorhombic $\beta$-FeSe$_{1-x}$ polymorph is superconducting with $T_c \sim 8$ K; neutron diffraction and Mössbauer spectroscopy show no evidence for long-range magnetic ordering, even though magnetic fluctuations on a shorter timescale were not ruled out [46,47,49,50]. A $^{77}$Se nuclear magnetic resonance analysis of $\beta$-FeSe$_{1-x}$ showed a strong enhancement of antiferromagnetic spin fluctuations with $\mathbf{k} \neq 0$ towards $T_c$, further increased by the application of hydrostatic pressure [55]. In contrast, first-principles electronic structure calculations suggest that the ground state of $\beta$-FeSe$_{1-x}$ should be in a collinear strip-like antiferromagnetic order [34]. Inelastic neuctron scattering measurements detected highly dispersive paramagnetic spin fluctuations with a strong magnetic response at $\mathbf{k} = (\pi,0)$ [56], thus revealing the presence of a fundamental component in the scenario foreseeing a pairing mechanism mediated by spin fluctuations.

Mismatching results are reported for the structural evolution of the $\beta$-FeSe$_{1-x}$ phase as a function of pressure. At room temperature a structural transformation takes place at $9 - 12$ GPa [57,58,59,60,61]. Laboratory X-ray diffraction analyses prompt to a $P4/nmm \rightarrow P6_3/mmc$ transition, giving rise to a hexagonal polymorph isotypic with NiAs [57,58]. Conversely synchrotron X-ray diffraction analyses indicate the occurrence of a $P4/nmm \rightarrow Pnma$ structural transformation [59,60,61], a result corroborated by theoretical calculations [61]. The $Pnma$ polymorphic modification is isotypic with FeAs and is still detected at low temperature ($8 - 16$ K) under high pressure. Uhoya *et al.* [60] and Kumar *et al.* [61] observed a $Cmme \rightarrow Pnma$ structural transformation around 10 K starting between 6.4 GPa and 9 GPa, completing at $\sim 31$ GPa. Margadonna *et al.* [62] report a 2-step structural transformation at 16 K: at 9 GPa the orthorhombic phase is partially transformed into the hexagonal polymorph, but further compression to $\sim 12$ GPa leads to the formation of the $Pnma$ phase. Then it is not clear whether the formation of the FeAs-type phase involves an intervening $\delta$-FeSe$_{1-x}$ phase; in fact, the same $\delta$-FeSe$_{1-x}$ phase transforms into the FeAs-type phase above $\sim 6$ GPa [62]. In any case, phase coexistence indicates that the transformation is 1$^{st}$ order. The $T_c$ increases up to $\sim 37$ K by applying a pressure of $\sim 7$-9 GPa, but further compression determines its progressive transformation into the $\delta$-FeSe$_{1-x}$ phase, with a subsequent decrease of $T_c$ [57,62]. Careful analyses [63,64] carried out under pressure up to 2.4 GPa showed that $\beta$-FeSe$_{1-x}$ is non-magnetic and that $T_c$ increases monotonically with increasing pressure for $p \lesssim 0.8$ GPa. For $0.8 \lesssim p \lesssim 1.2$ GPa, the superconducting and magnetic order parameters coexist, competing on a short length scale, and $T_c$ decreases with increasing pressure; a static, incommensurate magnetic order develops above $T_c$, but is partially/fully suppressed as superconductivity sets in. When $p \gtrsim 1.2$ GPa, the magnetic order is commensurate and coexists with superconductivity within the whole sample volume; this magnetic order is long-ranged, but the magnetic moment is very small [64].

### 2.1.3 The $Fe_{1+y}(Te_{1-x}Se_x)$ solid solution

Differing results are reported in some cases by the numerous studies on members of the β-$Fe_{1+y}$($Te_{1-x}Se_x$) solid solution. These differences can be largely ascribed to the fact that these materials are not stoichiometric and can contain different amounts of interstitial Fe; faint compositional variations can thus determine significant variances, especially with regard to magnetic properties. Therefore, in some cases, it becomes very challenging to compare data obtained from samples characterized by the same nominal composition, but prepared under different conditions, in particular when accurate structural and micro-structural analyses are lacking.

Evidence for a tendency towards phase separation is reported for several terms of the β-$Fe_{1+y}$($Te_{1-x}Se_x$) solid solution [65,66,67], even though this phenomenon could possibly be related to the cooling treatment after reacting annealing [68]. Samples with a nominal $x \sim 0.5$ equilibrated at 1073 K and 823 K were found to be constituted of two main compositions: $Fe(Te_{0.61}Se_{0.39})$ + $Fe(Te_{0.46}Se_{0.54})$ and $Fe(Te_{0.54}Se_{0.46})$ + $Fe(Te_{0.42}Se_{0.58})$, respectively [65]. We note that a miscibility gap in the pseudo-binary β-$FeSe_{1-x}$ - β-$Fe_{1+y}Te$ system cannot be stated by these analyses, since back-scattered scanning electron microscope images do not actually show any distinct interface that separates regions with different compositions [65]. Such a phenomenon can originate from the differences marking the crystal structures of the two end-members, implying a re-arrangement of both chemical bonds and geometrical interrelationships: 1) the strongly defective atomic plane in β-$Fe_{1+y}Te$ composed of interstitial Fe, which is absent in β-$FeSe_{1-x}$; 2) the significant differences in the chalcogen height displayed by β-$FeSe_{1-x}$ and β-$Fe_{1+y}Te$, determining a substantial compression along the $c$ axis of the tetrahedral layer in β-$FeSe_{1-x}$, absent in β-$Fe_{1+y}Te$. In the β-$Fe_{1+y}$($Te_{1-x}Se_x$) solid solution, the difference in the chalcogen height persists. In fact, both Se and Te ions are located at the same Wyckoff site $2c$, but do not have equivalent crystallographic orbits, since the variable coordinate $z$ values are significantly different, depending on the chalcogen atomic species [69,70,71]. This behaviour suggests a total lack of local structure relaxation. The chalcogen height disorder propagates to the Fe layer, where an elongation of the Fe thermal ellipsoid in the $c$-axis occurs [71].

The $P4/nmm \rightarrow Cmme$ structural transition characterizing the β-$FeSe_{1-x}$ phase is retained even after Te-substitution, up to the composition $Fe_{1.03}(Te_{0.43}Se_{0.57})$ [72]. Contrasting results are found for higher Te-contents: Li *et al.* [30] report a suppression of the structural transition in $Fe_{1.054}(Te_{0.507}Se_{0.493})$, but conversely, Bendele *et al.* [73] found an orthorhombic structure in a $Fe_{1.045}(Te_{0.594}Se_{0.406})$ sample at low temperatures. These differences can likely be ascribed to the different Fe content. In the terminal Te-rich solid solution, the $P4/nmm \rightarrow P2_1/m$ structural transition characterizing the $\beta$-$Fe_{1+y}Te$ phase is suppressed for $x \gtrsim 0.90$ [31,74,75].

The magnetic and superconductive properties in this system are strictly related to the crystallo-chemical features. In fact, magnetism is strongly affected by the Se content, the amount of interstitial Fe and the chalcogen height. The long-range antiferromagnetic ordering and the tetragonal-to-monoclinic structural transformation characterizing $\beta$-$Fe_{1+y}Te$ are suppressed in $Fe_{1+y}(Te_{1-x}Se_x)$ for $x \gtrsim 0.90$ [31,74,75], but short-range static magnetic ordering is retained up to $x \sim 0.45$ [76,77] with an incommensurate propagation vector $\mathbf{k} = (\frac{1}{2}\text{-}\delta,0,\frac{1}{2})$ [27,78]. This short-range magnetism has a static magnetic spin glass character [75,79]; such a phenomenon is likely related to the local structural disorder induced by the different chalcogen heights. The interstitial Fe favours static magnetic correlations with $\mathbf{k} = (\frac{1}{2}\text{-}\delta,0,\frac{1}{2})$, hence suppressing superconductivity when its amount exceeds the critical threshold [77,80,81]. Neutron diffraction measurements on $\beta$-$Fe_{1+y}(Te_{0.75}Se_{0.25})$ samples showed a broadening of the magnetic peak along $(\frac{1}{2}\text{-}\delta,0,\frac{1}{2})$ with the reduction of Fe content, suggesting a shortening of the magnetic correlations [82]. The chalcogen height determines the stability of the magnetic phases. In particular, density functional calculations indicate that for a $\beta$-$Fe_{1+y}Te$-type tetrahedron, in-plane $(\pi,0)$ spin fluctuations [in-plane wave vector $\mathbf{k} = (\frac{1}{2},0)$] dominate, favouring antiferromagnetism, whereas in a $\beta$-$FeSe_{1-x}$-type tetrahedron, in-plane $(\pi,\pi)$ fluctuations [in-plane wave vector $\mathbf{k} = (\frac{1}{2},\frac{1}{2})$] prevail, favouring superconductivity [33] (the maximum $T_c \sim 15$ K is found around $x \sim 0.5$ [73,83]). A rationale for this behaviour can be found in the shape of the Fermi surface. The mechanism of superconductivity in these systems is thought of in terms of antiferromagnetic spin fluctuations that are driven by Fermi surface nesting. Density functional calculations show that the Fermi surface nesting strongly varies with the chalcogen height (*i.e.* the distance between the Se/Te atom and the Fe plane), regardless of the chemical identity of the chalcogen atom [84]. This implies that the magnetic and superconducting properties are not directly affected by the chemical disorder. Se-substitution at Te sites reduces the chalcogen height, which in turns enhances the hybridization of the Fe-$3d$ states with the chalcogen-$p$ states. Fe-$3d$ bands across the Fermi levels widen, and the DOS at $E_F$ decreases, disfavouring the magnetic order. $Fe_{1+y}(Te_{1-x}Se_x)$ regardless the composition is characterized by disconnected Fermi surfaces consisting of hole sections around the zone centre (along $\Gamma$–Z), and two electron sections at the zone corner (along M –A direction). Holes Fermi surfaces are almost two-dimensional and with a circular sections. The holes Fermi surfaces shape is nearly unaffected by the chalcogen height.

Conversely, the electron Fermi surface changes in shape with the chalcogen height, with a pronounced three-dimensional character for high values of the chalcogen height and a more circular and two-dimensional character for lower values of the chalcogen height. The best nesting conditions is obtained when electrons Fermi surface is two-dimensional, a condition that is best achieved for $x \sim 0.5$, finding in good agreement with experiments showing that optimal doping in $Fe_{1+y}(Te_{1-x}Se_x)$ is about equal Te and Se concentrations. Interestingly for $x \geq 0.05$, dynamic magnetic correlations have been observed, characterized by an in-plane wave vector $\mathbf{k} = (\frac{1}{2} \pm \delta, \frac{1}{2} \mp \delta, l)$ corresponding to the same Fermi surface nesting characterizing the spin density wave state of the 122- and 1111-type compounds, and suggesting a common magnetic origin for superconductivity [74,85,86]. Hence, it has been argued that bulk superconductivity takes place as the static magnetic correlations with $\mathbf{k} = (\frac{1}{2}-\delta, 0, \frac{1}{2})$ are suppressed and those with an in-plane wave vector $\mathbf{k} = (\frac{1}{2} \pm \delta, \frac{1}{2} \mp \delta, l)$ become dominant, pointing to a strong correlation between superconductivity and the character of the magnetic order/fluctuations in this system [74,77]. In this scenario, non-bulk superconductivity arises before the complete suppression of long-range magnetism in the monoclinic structure [31,74]. There is not complete agreement about the exact Se content at which bulk superconductivity sets in. It is in some cases reported at $x \gtrsim 0.3$ [74,81], in other cases at $x \gtrsim 0.45$ [76,87]; these differences are probably related to different critical amounts of interstitial Fe. A different view was then argued after the analysis of oxidised single crystals: samples annealed in vacuum with $x \leq 0.3$ did not show bulk superconductivity; after air annealing, bulk superconductivity and antiferromagnetism were found to also coexist in the $0.05 \leq x \leq 0.18$ compositional range, indicating that $(\pi,0)$ and $(\pi,\pi)$ in-plane spin fluctuations can also coexist [88]. Air and $O_2$ annealing minimize the interstitial Fe content, thus suppressing $(\pi,0)$ in-plane spin fluctuations, and extending the bulk superconductivity field down to $x \sim 0.05 - 0.10$ [88,89,90]. It has not yet actually been clarified whether oxygen simply removes Fe-excess forming oxides at the surface [88,90], or whether some oxygen remains intercalated among the tetrahedral layers [89]; in the latter case, these kinds of samples should not strictly belong to the $\beta$-$Fe_{1+y}(Te_{1-x}Se_x)$ system. Moreover, an accurate structural characterization at low temperatures, which would ascertain whether the structure of these oxidized samples is tetragonal or orthorhombic, has not been found.

As for $\beta$-$FeSe_{1-x}$, an increase in $T_c$ (up to $\sim 23$ K) is also observed for the orthorhombic solid solution $Fe_{1.03}(Te_{0.43}Se_{0.57})$ by increasing the applied pressure up to $\sim 3$ GPa. As revealed by a structural analysis carried out at 14 K, around this pressure, the solid solution undergoes a $1^{st}$ order $Cmme \rightarrow P2_1/m$ transformation and further compression leads to a metallic, but not superconducting state [72].

At present, a complete phase diagram, covering in detail all of the afore-mentioned structural features of the pseudo-binary β-FeSe$_{1-x}$ - β-Fe$_{1+y}$Te system is not yet available. Nonetheless, a tentative pseudo-binary phase diagram can be assessed (Figure 2.2) by selecting data reported in the literature [46,30,31,72,73,74,75,76,83,81,87,91].

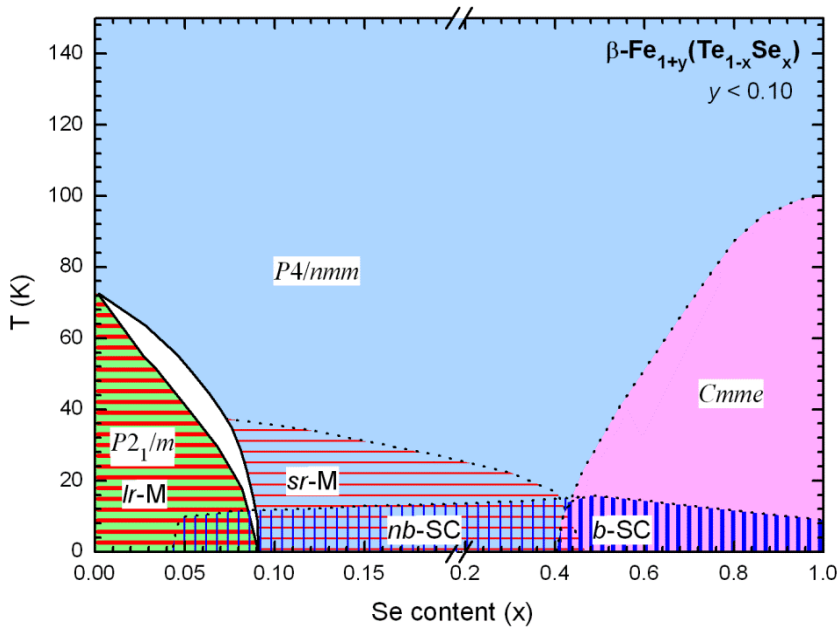

Figure 2.2: Assessed β-Fe$_{1+y}$(Te$_{1-x}$Se$_x$) phase diagram: lr-M: long-range ordered magnetic phase; sr-M: short-range ordered magnetic phase; b-SC: bulk superconducting phase; nb-SC: not-bulk superconducting phase; dotted lines represent approximate phase boundaries.

The reader should be aware that the definition of the phase boundaries depends to some extent on the content of interstitial Fe. In addition, different annealing treatments undergone by samples during synthesis often lead to slightly different transition temperatures, complicating to some extent the exact definition of the equilibrium curves; the phase diagram must however represent the thermodynamic equilibria. Understanding the relationship between the orthorhombic structure and bulk superconductivity is key; unfortunately, most papers describing samples that exhibit bulk superconductivity lack an accurate structural characterization at low temperatures. In this context, a sample with a Fe$_{1.045}$(Te$_{0.594}$Se$_{0.406}$) measured composition was found to be orthorhombic with bulk superconductivity [73]; conversely, superconducting samples with Fe$_{1.088}$(Te$_{0.584}$Se$_{0.416}$) and Fe$_{1.054}$(Te$_{0.507}$Se$_{0.493}$) compositions were found to be tetragonal at low temperatures, but with in-plane $(\pi,0)$ spin fluctuations [30], suggesting non-bulk superconductivity. These structural differences can likely be ascribed to different contents of interstitial Fe; as a consequence, in the phase diagram of Figure 2.2, bulk superconductivity was plotted within the orthorhombic field.

## 3. The 122-type systems

The sub-class of 122-type compounds is basically constituted of $CaFe_2As_2$, $SrFe_2As_2$, $BaFe_2As_2$, $EuFe_2As_2$ and their derivatives. They crystallize into the tetragonal $I4/mmm$ - 139 space group at room temperature and are isotypic with tetragonal $ThCr_2Si_2$. During cooling, a $I4/mmm \rightarrow Fmmm$ structural transformation takes place, coupled with a magnetic transition [92], leading to an orthorhombic β-$ThCr_2Si_2$ structure-type. The study of this structural transformation would already be important for understanding the concurrent magnetic transition, but the discovery of the isotope effect in superconductivity suggests that structural effects play a role in the understanding of superconductivity [93]. The nature of these transitions as well as their relationship have long been debated [94]; two scenarios have been proposed, the first one involving a single magneto-structural transition; the second one involving a proper or pseudo-proper ferroelastic transformation preempting the magnetic transition. Contrasting results are reported for the order of the structural transformation: a $2^{nd}$ order nature was argued for $SrFe_2As_2$ ($T_s = T_m \sim 200$ K) (despite the occurrence of hysteresis and a volume discontinuity) [95], and $BaFe_2As_2$ ($T_s = T_m \sim 140$ K) [96,97]. Subsequent studies definitively ascertained that in $CaFe_2As_2$ ($T_s = T_m \sim 170$ K), $SrFe_2As_2$ and $BaFe_2As_2$, the structural transformation is $1^{st}$ order [98,99,100,101,102,103].

The concomitant magnetic transition is reported to be $1^{st}$ or $2^{nd}$ order [102,97] in $BaFe_2As_2$, but $1^{st}$ order in $SrFe_2As_2$ [101] and $CaFe_2As_2$ [103]. The magnetic structure belongs to the $F_Cmm'm'$ Opechowski-Guccione notation for magnetic space group [94,104,105] (or $C_Amca$, according to the Belov-Neronova-Smirnova notation [106]), and is characterized by an antiferromagnetic spin ordering whose propagation wave-vector is $\mathbf{k} = (1,0,1)$ in an orthorhombic unit cell with $c > a > b$; the ordered Fe moment never exceeds 1 $\mu_B$. Spin ordering results in a stripe-like structure, with ordered Fe-moments oriented along the $a$-axis in the $a$-$b$ plane; antiferromagnetic spin coupling occurs along the $c$-axis and the longer orthorhombic $a$-axis, whereas along the shorter orthorhombic $b$-axis, ferromagnetic coupling is present (Figure 1.4). The $EuFe_2As_2$ compound exhibits another magnetic transition at $\sim 18$ K, which is associated with the spin ordering of the magnetic $Eu^{2+}$ ions [107]. Upon application of pressure, the magnetic transition is suppressed and the undoped 122-type compounds become superconducting [108,109,110,111,112,113].

The parent $BaFe_2As_2$ composition lead to the most studies, because structurally simple and large crystals can be grown; in addition, in the Ba-based 122 systems, the systematic substitution of Ba, Fe or As atom with a different element can in several cases drive the antiferromagnetic state of the parent compound to a superconducting ground state. A very careful study revealed that magnetic ordering occurs $\sim 0.75$ K below the orthorhombic distortion [114]. In addition, the structural transformation exhibits a very peculiar behaviour: two different orthorhombic phases coexist within

~ 1 K around $T_s$ [114,115]. In particular, the former phase is paramagnetic and characterized by a reduced orthorhombic distortion, and is rapidly suppressed below $T_s$, whereas the latter phase is antiferromagnetic and stable down to the lowest temperature [114]. This behaviour likely shows the dependence of structural properties on the magnetic ordering developing within the paramagnetic low-distorted orthorhombic phase; as the magnetic order percolates in the paramagnetic orthorhombic phase, an effective shear stress occurs, increasing the lattice distortion in the magnetically-ordered phase. An unusual biquadratic coupling between the structural and magnetic order parameters has been observed, suggesting the proximity of the system to a tetracritical point [97,104]; this kind of coupling possibly indicates that the structural distortion is driven by an independent ferroelastic instability, rather than by the magnetic ordering [94].

Since the ionic radius of $Ca^{2+}$ is notably smaller than those of both $Sr^{2+}$ and $Ba^{2+}$, a significant shortening of the $c$-axis occurs in the $CaFe_2As_2$ phase. By application of external pressure, an isomorphous $I4/mmm \rightarrow I4/mmm$ structural transformation takes place, leading to the formation of a collapsed tetragonal structure, inside which As-As chemical bonds and a concomitant suppression of magnetic ordering occur. At low temperatures, this collapsed phase becomes superconducting [116,117,118]. Later the collapsed phase was also detected under high pressure in $BaFe_2As_2$ [119], $SrFe_2As_2$ [120] and $EuFe_2As_2$ [121].

The microscopic origin of the structural transition is not clear. A theoretical study based on the Ginzburg-Landau approach [122] gives us some insights on the interplay between structural and magnetic phase. In this work, the authors provide a unified framework that explains the different experimental findings, notably the simultaneity of the structural and magnetic transitions, and their character as $1^{st}$ or $2^{nd}$ order transitions. The presence of magneto-elastic coupling is at the origin of the two transitions. By minimizing the free energy of the system, a quadratic dependence of the strain on the magnetization is obtained. This does not allow the existence of the magnetic transition alone, that is, at a temperature higher than that of the structural one ($T_s < T_m$). Therefore, even if the origin of the two transitions is not clear, we know that the magnetic order alone is able to drive the structural distortion, which is exactly what happens in the materials where $T_s = T_m$. If the structural transition comes first ($T_s > T_m$), it has to originate from a different source. This may be a genuine ferroelastic transition originating from the vanishing of the elastic modulus or from the effect of a spin nematic order. In the nematic order, the spins time-averaged value on each Fe atom is zero because of the fluctuations, but the instantaneous ordering of the magnetic moments is anti-ferromagnetic. The nematic order is able to drive the structural transition above the Neel temperature, and it may be the mechanism behind the higher value of the critical temperature of structural transitions. It is suggested that a criterion to find which of the two mechanisms (ferroelasticity or nematic order) are at the origin of the structural transition is given by the behaviour of the elastic constant, which is linear with temperature for the ferroelastic transition, with a square-root dependence on temperature for the nematic order. A recent experimental study of elastic moduli in hole and electron-doped $BaFe_2As_2$ confirms the square-root behaviour for the elastic modulus, supporting the existence of a nematic order [123]. The origin of the nematic order is not clear. Nematic order can arise as the effect of orbital order, or may be originated by a genuine magnetic transition that lowers the system symmetry (the so-called Ising-nematic phase). Unfortunately, both orbital order and Ising-nematic order break the same symmetry, which makes it hard to distinguish one order from the other from experimental evidence. In Ref. [4], a microscopic model based on itinerant electrons is used to shed light on the origin of the structural transition. The authors show that the stripe magnetic order is generally preempted by an Ising-nematic order. The nematic transition may instantly bring the system to the verge of a magnetic transition, or it may occur first, being followed by a magnetic transition at a lower temperature. Furthermore, due to the distinct orbital character of each Fermi pocket, the nematic transition also induces orbital order.

Superconductivity is commonly induced by electron- or hole-doping. Electron doping in 122-type compounds is usually obtained by *TM*-substitution (*TM* = Co, Ni, Rh, Pd, Ir, Pt); in all of these cases, the phase diagrams are almost coincident after appropriate scaling. Remarkably, when the substituting elements belong to a same chemical group, as in the case of Co and Rh or Ni and Pd, the corresponding phase diagrams are amazingly almost exactly coincident [124].The *Ln*-substitution constitutes an alternative, but uncommon mode, to gain electron-doping, requiring a high pressure synthesis method [125]. We note that the highest $T_c$ in 122-type compounds (up to 45 K) has been measured in the collapsed phase of $(Ca_{1-x}Ln_x)Fe_2As_2$ compounds (*Ln* = La - Nd) [126]. Superconductivity by hole-doping is usually achieved by *A*-substitution (*A* = Na, K), suppressing the antiferromagnetic ordering, but not by *TM*-substitution (*TM* = Mn, Cr). Interestingly, magnetism can be suppressed and superconductivity can also be achieved by isovalent doping (that is chemical substitution of Ru and P at the Fe and P site), respectively.

### 3.1 Substitution and doping of the BaFe₂As₂ phase

$BaFe_2As_2$ can be considered the prototypical phase of the 122-type compounds; in fact, this phase is particularly suitable for a systematic study of the dependence of the structural, magnetic and superconductive properties on the charge carrier density, since it can either be electron- or hole-doped by partial Co- or K-substitution, respectively (Figure 3.1).

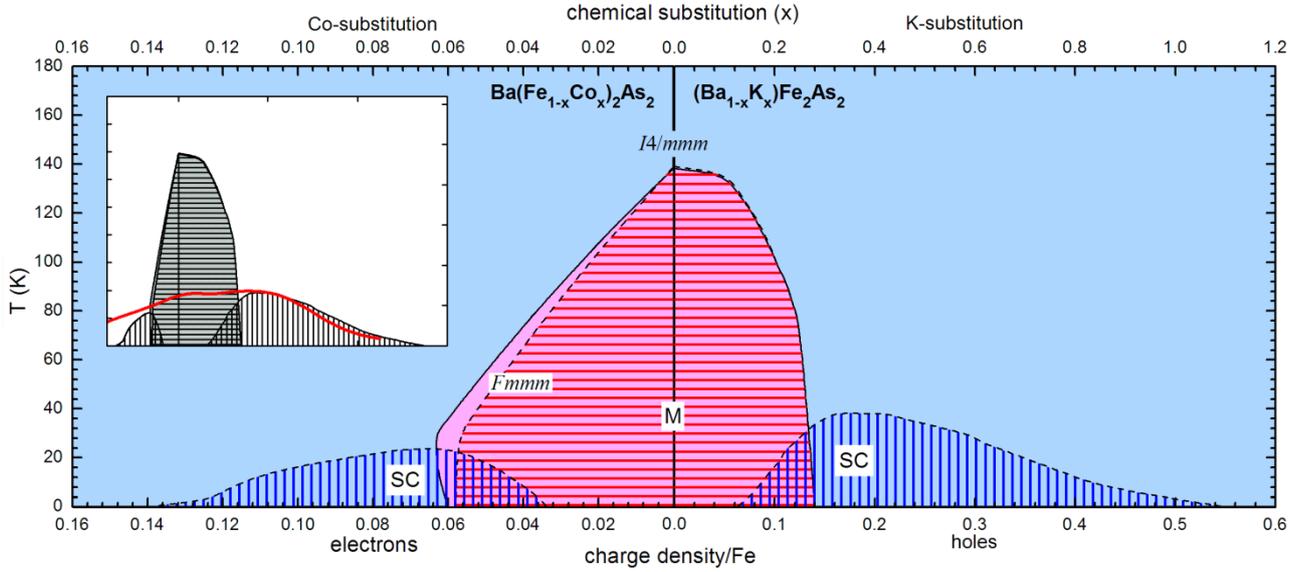

Figure 3.1: Evolution of the phase diagram of BaFe$_2$As$_2$ as a function of the charge density, from the electron-doped (Co-substitution) to the hole-doped (K-substitution) regime; data are taken from ref. [104,130,132,136,127,128,146]; in order to better highlight the features characterizing the Ba(Fe$_{1-x}$Co$_x$)$_2$As$_2$ system, the $x$-scales at the electron- and hole-doped sides are different. The inset shows the actual proportions among the different phase fields; the red curve represents the doping dependence of the Lindhard function at the M point, as calculated in ref. [138].

The phase diagrams of the (Ba$_{1-x}$K$_x$)Fe$_2$As$_2$ and Ba(Fe$_{1-x}$Co$_x$)$_2$As$_2$ systems are quite similar and the main differences concern the superconductive field. In the Ba(Fe$_{1-x}$Co$_x$)$_2$As$_2$ system, the superconductive dome is narrower with the highest $T_c \sim 25$ K, against the $\sim 38$ K of the (Ba$_{1-x}$K$_x$)Fe$_2$As$_2$ system. In particular, the superconducting dome extends to much higher doping in the hole-doped side, with optimal concentrations of $\sim 0.17$ hole/Fe $vs \sim 0.06$-$0.07$ electron/Fe in the opposite-doped sides. In the under-doped regions, both systems are characterized by a microscopic phase coexistence between the superconductive state and the antiferromagnetic ordering in the orthorhombic phase, which is suppressed as the optimal doping is approached [104,129,130,131,132,133,134]. In this context, it is worth to note that, conversely, in the Ru-substituted samples, the magnetic and superconductive states coexist in a not homogenous way [135].

In the (Ba$_{1-x}$K$_x$)Fe$_2$As$_2$ system, the structural and magnetic transitions are 1$^{st}$ order and coincident [130], whereas in the Ba(Fe$_{1-x}$Co$_x$)$_2$As$_2$ system, the structural transformation and magnetic transition are split and 2$^{nd}$ order, except at very small Co-concentration [114,136]. The interpretation of the coupled structural and magnetic transitions based on Ginzburg-Landau theory [122] provides an explanation of this behaviour.; a magnetic transition alone would be a 2$^{nd}$ order one, but the magneto-elastic coupling with the system structure turns the transition to 1$^{st}$ order one whenever $T_m$ and $T_s$ are coincident or very close.

The electron-hole asymmetry of the phase diagram can be understood by inspecting the band structure of $BaFe_2As_2$. Like most of the iron-pnictides, $BaFe_2As_2$ is a compensated semimetal, that is, valence and conduction bands overlap at the Fermi level forming holes and electron pockets in different locations of the first Brillouin zone. The compensation of the semimetal comes from the electron counting that satisfies the octet rule, so that iron-pnictides would be insulators if their bands were not overlapping. According to the Luttinger theorem, electron and hole pockets have identical volumes in the undoped compound (e.g. $Ba\ Fe_2As_2$). Theoretical calculations [137] show that the shape of Fermi surfaces and the effective masses of the electron and holes are different. Therefore, an asymmetric behaviour of the phase diagram is justified on the basis that the Fermi nesting condition will be different if electron or holes doping is used. This is also confirmed by the calculation of the Lindhard spin susceptibility. The doping dependence of the Lindhard spin susceptibility at the M point roughly reproduces the asymmetry between the electron- and hole-sides (Figure 3.1, inset), supporting a scenario where superconductivity is driven by a Fermi surface quasi-nesting [138]. We note that the maximum value of the calculated susceptibility results near the experimental optimal hole doping and the Lindhard function tracks the superconducting transition qualitatively well in the hole-doped side [138]. Conversely, in the electron-doped side, the highest $T_c$ and the extension of the superconducting field are over-estimated [138]; this behaviour likely originates from the structural disorder produced by Co-substitution at the Fe sub-lattice. As for the problem of the symmetry of the superconducting pairing state, two candidates for the order have been suggested: (i) $s^{+-}$, an unconventional state in which the sign of the superconductive order parameter is opposite on the electron and hole Fermi surface pockets (ii) $s^{++}$, a conventional state where the sign of the order parameter is the same on both pockets. Clues on this issue can be obtained by a theoretical study within the framework of weak mean-field theory: a simple model that makes use of circular (elliptical) hole (electron) Fermi surface with constant magnetic and pairing interactions is able to predict that only the unconventional $s^{+-}$ superconducting pairing state is compatible with the experimental evidence of a coexisting itinerant magnetism and superconducting state [139,140]. Moreover, the excitation spectrum of the $s^{+-}$, is predicted to be gapped. Instead, for the conventional $s^{++}$ state, the coexistence of magnetism and superconductivity is ruled out. Furthermore, the suppression of $T_c$ in the under-doped regime results from the competition between magnetism and superconductivity, whereas in the over-doped region it originates from the change in the Fermi surface with doping [140].

Experimental evidence for the coexistence between magnetic and superconducting order in the electron doped $Ba(Fe_{1-x}Co_x)_2As_2$ compound, supports the existence of the $s^{+-}$ superconducting order [133,141].

### 3.1.1 The Ba(Fe$_{1-x}$Co$_x$)$_2$As$_2$ system

In the Ba(Fe$_{1-x}$Co$_x$)$_2$As$_2$ system, the structural transformation and magnetic transition are split; the structural transformation is 2$^{nd}$ order, whereas the magnetic transition changes from 1$^{st}$ to 2$^{nd}$ order at the tricritical point $x \sim 0.022$ [114,136]. Several experiments revealed that superconductivity and antiferromagnetism compete and microscopically coexist in the under-doped regime [131,142], similarly to what has been observed in other electron doped compounds such as Ba(Fe$_{0.9625}$Ni$_{0.0375}$)$_2$As$_2$ [143] and Ba(Fe$_{0.961}$Rh$_{0.039}$)$_2$As$_2$ [144].

In the $0.035 \leq x \leq 0.063$ compositional range, superconductivity coexists with long-range antiferromagnetism in the low-temperature orthorhombic phase, but the orthorhombic distortion decreases during cooling as the superconductive phase field is entered. We note that magnetism becomes incommensurate for $x > 0.56$, a phenomenon that is consistent with the formation of a spin-density wave [145]; for $x \geq 0.66$, the structural transformation leading to the formation of the orthorhombic phase is definitively suppressed [146]. Such behaviour was ascribed to a competition for the same electronic state between two electronically driven orders: the superconductive and nematic (orthorhombic) states, suggesting an electronic character of the nematic transition [20,139,147]. Theoretical analyses based on the Ginzburg-Landau approach show that the possible scenarios for the superconducting and magnetic phase coexistence are limited, but two cases are possible [140]. In the first one, named homogeneous coexistence, the competition between superconductivity and magnetism leads to the genuine coexistence of both phases. The homogeneous coexistence is characterized by a tetracritical point where the two-phase lines (superconducting and magnetic) cross. Below the tetracritical point, a region of homogeneous coexistence is found. The region is surrounded by two 2$^{nd}$ order phase lines. Moreover, the shape of the phase lines is influenced by competition among the phases. Indeed, superconductivity tends to suppress magnetism as the temperature goes to zero, leading to the decrease in the orthorhombic distortion and the re-entrance of the nonmagnetic phase. This scenario is consistent with experimental observations, where the coexistence of superconductivity and magnetism at an atomic scale is observed [133,141]. The second possible scenario, named heterogeneous coexistence, does not allow for a real coexistence of the two phases and is thus not representative of the actual features of the phase diagram. In this case, superconducting order and magnetism are segregated in non-overlapping regions of space. Here, the cross point between the phase lines is called bicritical, and the coexistence region is surrounded by first-order phase lines.

We note that, in this system, the nematic state in the tetragonal phase extends above both the magnetic ordered field and the entire superconducting dome [148].

### 3.1.2 The (Ba₁₋ₓKₓ)Fe₂As₂ system

Hole doping in BaFe$_2$As$_2$ can be obtained by K-substitution; the (Ba$_{1-x}$K$_x$)Fe$_2$As$_2$ system has been extensively studied, since the highest $T_c$ (up to ~ 38 K) for 122-type compounds is attained in optimally-doped samples ($x$ ~ 0.4), where both structural and magnetic transitions are suppressed [92,149]. In this system, the BaFe$_2$As$_2$ and KFe$_2$As$_2$ end-members are isostructural, and the structural transformation and magnetic transition are simultaneous and 1$^{st}$ order [104]. We note that in slightly under-doped (Ba$_{1-x}$K$_x$)Fe$_2$As$_2$ samples ($T_c$ = 32 K), the magnetic order sets in without symmetry breaking below $T_m$ = 70 K [150]. As a consequence, the tetragonal symmetry is preserved, but the crystal structure undergoes an increase in the lattice micro-strain without a macroscopic breakdown of the lattice symmetry [150]. The structural order parameter decreases as the phase enters the superconductive phase field, but the structure remains orthorhombic, conversely to what observed in the electron-doped Ba(Fe$_{1-x}$Co$_x$)$_2$As$_2$ and Ba(Fe$_{1-x}$Rh$_x$)$_2$As$_2$ systems [129,144], where a re-entrance occurs. Such a different behaviour can probably be ascribed to the structural disorder induced by Co and Rh atoms at the Fe sub-lattice, absent in the (Ba$_{1-x}$K$_x$)Fe$_2$As$_2$ system. The biquadratic coupling between the structural and magnetic order parameters observed in pure BaFe$_2$As$_2$ apparently persists over a wide compositional range in the (Ba$_{1-x}$K$_x$)Fe$_2$As$_2$ system; it is possible that this apparent behaviour is a trick, whereas a linear-quadratic coupling is actually present [104].

### 3.1.3 The (Ba₁₋ₓKₓ)(Fe₁₋ᵧCoᵧ)₂As₂ system

Very interesting clues can be gained by studying the charge-compensated (Ba$_{1-x}$K$_x$)(Fe$_{1-y}$Co$_y$)$_2$As$_2$ system ($x/2$ ~ $y$) [151]. For $y \leq 0.13$, the orthorhombic phase and the antiferromagnetic ordering are stable at low temperatures, but the structural and magnetic order parameters are reduced by increasing the degree of substitution. Significant magnetism persists up to $y$ = 0.19, whereas the orthorhombic distortion is detected only up to $y$ = 0.13. For $0.15 \leq y \leq 0.19$, bulk superconductivity (highest $T_c$ ~ 15 K) coexists with a static magnetic order on a microscopic scale within a tetragonal structure. As the substitution level exceeds $y \geq 0.25$, a tetragonal non-magnetic state takes place. From the observed linear relationship between the structural and magnetic order parameters, it was concluded that the orthorhombic and superconductive phases are both controlled by magnetic instability [151].

In the (Ba$_{1-x}$K$_x$)(Fe$_{1.86}$Co$_{0.14}$)$_2$As$_2$ system, the electron doping induced by a fixed amount of Co-substitution was progressively compensated by hole-doping with K-substitution [152]. The orthorhombic distortion and the magnetic transition are both recovered at low levels of hole-doping.

With the increase of K-substitution, hole- and electron-doping are perfectly compensated and superconductivity is completely suppressed, as in the pure parent compound $BaFe_2As_2$. Further increase of K-content moves the system within the hole-doped region of the system and recovers superconductivity with the highest $T_c \sim 30$ K. This critical temperature is lower than those measured in the $(Ba_{1-x}K_x)Fe_2As_2$ system, probably on account of the disorder in the Fe sub-lattice produced by Co-substitution.

### 3.1.4 The $(Ba_{1-x}Na_x)Fe_2As_2$ system

The magnetic and superconducting properties exhibited by the $(Ba_{1-x}Na_x)Fe_2As_2$ system are quite similar to those of the homologous $(Ba_{1-x}K_x)Fe_2As_2$ system; despite the large mismatch between the $Ba^{2+}$ and $Na^+$ ionic radii, no evidence for ordering was observed [153]. The structural and magnetic phase transitions are coincident and both 1st order [105]. A first systematic study concluded that the tetragonal-to-orthorhombic transformation occurs during cooling up to $x = 0.35$, while for larger Na-content, the tetragonal $I4/mmm$ is retained down to the lowest temperature [154]. Subsequently, a peculiar magnetic phase was detected in the compositional range $0.24 \leq x \leq 0.28$ [155], where superconductivity coexists with magnetism at low temperatures. More interestingly, these compositions were found to first undergo a tetragonal-to-orthorhombic structural transformation during cooling; in a second stage, as the temperature is further decreased, a spin re-orientation occurs and the orthorhombic phase becomes unstable. As a consequence, a 1st order transition takes place and the orthorhombic phase is partly transformed into the pristine tetragonal one; both phases display antiferromagnetic ordering and appear to be superconductive [105,155]. The structure of the low-temperature tetragonal phase is stretched in the *ab*-plane and compressed along the *c*-axis, whereas the magnetic structure of the orthorhombic phase does not qualitatively change across the transition [156]. The magnetic structure associated with the tetragonal phase is described by an antiferromagnetic stripe model with moments that are polarized along the *c*-axis [156]. The re-entrant magnetostructural transitions characterizing this system were analyzed for both magnetically and orbitally-driven mechanisms, but at present, the underlying physical mechanism has not yet been unveiled [156].

These results clearly indicate that magnetism competes with superconductivity; for this reason, it was first suggested that the nematic order is possibly of magnetic origin [155], but a subsequent analysis concluded that the structural transition has a purely electronic origin [156].

In any case, the occurrence of a magnetic phase associated with the tetragonal structure, localized only within a two-phase field, raises some concerns: it is in fact not clear why it is completely suppressed as the orthorhombic phase disappears. The strongly strained nature of the tetragonal phase in the bi-phasic field likely plays a dominant role. We note that in the phase diagrams drawn in ref. [105,155], the phase field pertaining to the so-called $C_4$ phase is actually a biphasic field, in which the orthorhombic and tetragonal structures coexist [105,155]. In this system, the two-phase field is likely very narrow above ~ 50 K, but broadens as the temperature is further cooled. This gives rise to the rather unusually extended separation between the tetragonal and the orthorhombic phase fields at low temperatures, which is not observed in the other 122-type systems.

Another question concerns the homogenous variation of $T_c$ with composition throughout the orthorhombic-to-tetragonal phase field. In fact, no net discontinuity is observed at the structural transition in all of the reported phase diagrams for 122- and 1111-type systems. This phenomenon becomes extraordinarily apparent in the $(Ba_{1-x}Na_x)Fe_2As_2$ phase diagrams of ref. [105,155], in which this homogenous variation crosses a relatively wide and peculiar two-phase field.

The re-entrant structural transformation, demixing and the tetragonal magnetic phase were not confirmed by a new study based on neutron diffraction analyses of single-crystal samples [157]; in this case, the analytical results pointed to a spin reorientation along the $c$-axis in the orthorhombic magnetic phase, inducing structural changes in the orthorhombic crystalline structure itself. As a result, spin re-orientation appears as the characterizing feature of the low-temperature transition, whereas no evidence for a coexisting tetragonal phase was found [157].

### 3.1.5 The BaFe₂(As₁₋ₓPₓ)₂ system

In the $BaFe_2(As_{1-x}P_x)_2$ system, the isovalent substitution suppresses both the structural transition and the spin density wave, inducing superconductivity up to 30 K [158]. In this system the tetragonal to orthorhombic transformation and magnetic ordering occur concurrently as a 1st order transition [159]. The existence of a quantum critical point in this system at $x \sim 0.3$ is a conundrum; former analyses suggested that the phase transition from the coexistence field to the superconducting state is 2nd order, corroborating its occurrence [158,160]. Later, antiferromagnetism and superconductivity was found to coexist and compete, pointing to a scenario with an avoided quantum critical point [161]. Moreover some investigations estimated the suppression of the orthorhombic phase around $x \sim 0.28$ [159], although in a recent single crystal synchrotron X-ray diffraction investigation a re-entrant behaviour of the orthorhombic structure in the temperature range 26 K $\leq T \leq$ 32.5 K has been observed in a sample with $x = 0.28$ [161].

The role of orbital ordering in this system has been highlighted by [75]As-NMR analysis [162]; by means of these measurements the orbital polarization of the As $4s$ orbitals, likely originated by orbital ordering of the Fe $3d$ orbitals, has been detected within electronic domains pertaining to the tetragonal phase. More interestingly this polarization is found already static near room temperature, suggesting that fluctuating orbital order can be pinned by the substituting P atoms, acting as structural defects.

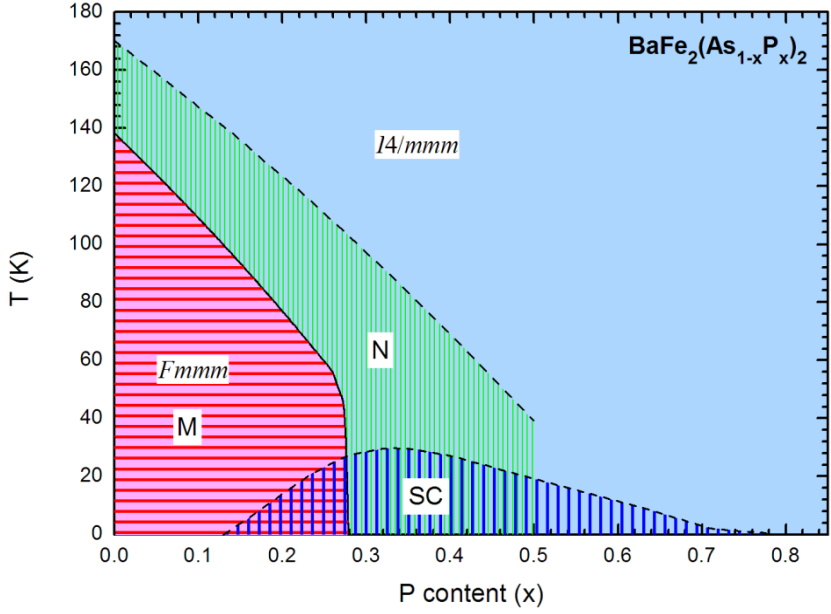

Figure 3.2: The phase diagram of the $BaFe_2(As_{1-x}P_x)_2$ system; data from ref. [158,160,164,165].

As for the β-$Fe_{1+y}(Te_{1-x}Se_x)$ solid solution, both As and P ions have different orbits; they are located at the same Wyckoff site 4$e$, but display significantly different variable coordinate $z$ values; this feature determines slightly different Fe-As and Fe-P bond lengths, likely suppressing both the structural and magnetic transitions, and inducing superconductivity. Conversely, in the homologous $(Ba_{1-x}Sr_x)Fe_2As_2$ system, where the Fe-As bond lengths are constant, both transitions are retained [163]. We note that, the phase diagram of the $BaFe_2(As_{1-x}P_x)_2$ system is qualitatively similar to that of the $Ba(Fe_{1-x}Co_x)_2As_2$ system.

This system is particularly interesting because a systematic determination of the nematic transition temperature ($T_n$) as a function of composition was carried out [164]. We note that the analysis of selected Bragg peaks obtained by synchrotron X-ray diffraction showed a line broadening coupled with a suppression of the relative intensity, even for the optimally substituted composition ($x$ = 0.33; $T_c$ = 31 K). Such a behaviour, which in the under-substituted samples precedes symmetry breaking, was ascribed to the formation of the electronic nematic state with $T_n$ = 85 K. These authors concluded that at $T_n$ a true phase transition occurs, where the $C_4$ rotational symmetry is broken, whereas at $T_s$ a meta-nematic transition is present [164]. Within this scenario the orthorhombic phase establishes at much higher temperature and extends over (almost) the whole superconducting phase field.

Using data reported in the literature [158,160,164,165], a complete phase diagram showing the stability fields of the nematic, orthorhombic, magnetic and superconductive phases can be drawn for the $BaFe_2(As_{1-x}P_x)_2$ system (Figure 3.2). We note that the nematic phase occurs even in the superconductive over-doped regime, where long-range magnetic ordering is absent, similarly to what is observed in the $Ba(Fe_{1-x}Co_x)_2As_2$ system [148]. This phenomenon contrasts with theoretical models foreseeing that magnetic ordering is generally pre-empted by an Ising-nematic transition, which also induces orbital ordering [4], and possibly indicates that orbital physics plays a primary role.

### 3.2 Substitution and doping of the CaFe₂As₂ phase

The structural and physical properties displayed by the substituted $CaFe_2As_2$ phase are, in many cases, notably different from those of the homologous substituted $BaFe_2As_2$ systems. Nonetheless upon application of pressure the $CaFe_2As_2$ phase becomes superconducting, similarly to other undoped 122-type compounds; the nature of the superconducting phase in $CaFe_2As_2$ has been deeply discussed, but later specific heat measurements ascertained pressure induced bulk superconductivity below 7 K [166].

Conversely in the substituted systems bulk superconductivity has not yet firmly confirmed. In the electron doped $Ca(Fe_{1-x}Co_x)_2As_2$ system, superconductivity abruptly arises and then progressively decreases with further doping [167]. In the $Ca(Fe_{1-x}Rh_x)_2As_2$ system, the usual tetragonal-to-orthorhombic transformation is observed up to $x \sim 0.20$; further substitution induces superconductivity, but for $x \geq 0.24$, a non-superconducting collapsed tetragonal phase takes place [168]. Finally, the $CaFe_2(As_{1-x}P_x)_2$ system exhibits a 1[st] order isomorphous $I4/mmm \rightarrow I4/mmm$ structural transformation for $x > 0.05$, leading to the formation of a collapsed tetragonal phase [169], similarly to what is observed in the pure $CaFe_2As_2$ phase under pressure. This collapsed phase coexists with the tetragonal phase within a relatively wide range of compositions [169], indicating that the transition along the chemical composition axis is also 1[st] order.

## 4. The 1111-type systems

There exist two prototypical compositions for the 1111-type compounds: $Ln$FeAsO ($Ln$: lanthanide) and $AE$FeAsF ($AE$: alkaline earth), the first one being the most extensively studied. These compounds crystallize at room temperature in the $P4/nmm$ - 129 space group, isotypic with ZrCuSiAs. Un-substituted compounds undergo a 1[st] order $P4/nmm \rightarrow Cmme$ structural transformation at $T_s$, ranging around 140 – 180 K. The structural transformation is followed by a 2[nd] order magnetic transition at $T_m$ [170], a few tens of degrees below $T_s$, involving spin ordering at the Fe sub-lattice. In $AE$FeAsF compounds, the decoupling is larger, around $\sim$ 50 K as in SrFeAsF [171].

$T_m$ depends slightly on the $Ln$ atomic species, ranging from ~ 130 K to ~ 150 K, but with the decrease of the $Ln^{3+}$ ionic radius, both $T_s$ and $T_m$ decrease and tend to converge [172]. The ordered Fe moment is always lower than 1 $\mu_B$, as for the 122-type materials. The magnetic structure is characterized by an in-plane wave vector $\mathbf{k}$ = (1,0) in orthorhombic notation; depending on the $Ln$ atomic species, different couplings are observed along the $c$-axis [9]. The magnetic moments lie in the $ab$-plane, along the longer $a$-axis [173]. A second magnetic transition that takes place at lower temperatures in several cases, is originated by the antiferromagnetic ordering of magnetic $Ln^{3+}$ ions. Electron or hole doping is required in order to suppress static magnetic ordering and induce superconductivity. There are three main ways to induce electron doping in $Ln$FeAsO compounds: 1) partial substitution of O with F; 2) partial substitution of Fe with Co (or other electron-richer elements such as Ir and Ni); 3) electron doping with H⁻. Hole doping can be achieved by 1) partially replacing $Ln^{3+}$ with $AE^{2+}$ ions or 2) introducing vacancies at the O sub-lattice; in $AE$FeAsF compounds hole doping is obtained by partial substitution of F with O.

In some cases, superconductivity can also be achieved by isovalent doping, as P-substitution in LaFeAsO [174] SmFeAsO [175], suppressing both structural and magnetic transitions. The CeFe(As$_{1-x}$P$_x$)O system exhibits a rather peculiar behaviour: initial studies failed to detect the superconductive state [176,177], which was subsequently determined to be within a narrow homogeneity range around $x$ ~ 0.30 [178]. Conversely, Ru-substitution does not induce superconductivity in any case [179,180].

Some un-substituted compounds become superconducting under high pressure, such as LaFeAsO [181,182] and SmFeAsO [182], but not CeFeAsO [183].

The La-based systems, LaFeAs(O$_{1-x}$F$_x$) and LaFeAs(O$_{1-x}\square_x$), are characterized by the lowest $T_c$ among the 1111-type family, but by applying an external pressure of 4 GPa to optimally doped LaFeAs(O$_{1-x}$F$_x$) compounds, the $T_c$ has been raised up to 43 K [184], whereas in LaFeAs(O$_{1-x}\square_x$), a superconducting $T_c$ onset of ~ 50 K has been measured under 1.5 GPa [185]. On the other hand, $T_c$ can also be increased in these systems by chemical pressure, by partial substitution of La with Y or Sm [186,187,188]. The complete replacement of La with other $Ln$ elements, such as Sm, Ce, Nd, Pr, and Gd, increases $T_c$ up to ~ 55 K (at present, maximum $T_c$ = 58.1 K for a Fe-based superconductor has been measured in SmFeAs(O$_{0.74}$F$_{0.26}$) [189]). We note that $T_c$ decreases by applying pressure to these high- $T_c$ systems [190].

As in other Fe-SC, the relationship between the magnetic and superconductive phases is one of the most studied topics. Two different behaviours are reported for $Ln$FeAsO systems at the verge of the magnetic - superconducting phase boundary: 1) mutual exclusion between the magnetic and superconducting phases; 2) microscopic phase coexistence. At present, it is not clear whether a general behaviour holds, since controversial results are also reported for the same system; more accurate analyses are needed in order to determine whether the involved *Ln* ion plays a primary role in controlling the electronic-magnetic phase equilibria.

## 4.1 Electron doped systems: F-substitution

F-substitution is the most commonly applied method for inducing superconductivity in 1111-type systems, since highest $T_c$ are obtained; optimal doping is generally achieved for $x \sim 0.15$. Such a substitution progressively decreases the orthorhombic distortion of the lattice and hinders Fe-magnetism, which is always observed within the orthorhombic phase. It is generally stated that above a critical F-content $(x_c)$, the tetragonal-to-orthorhombic transformation is completely suppressed; Table 2.4.I lists the maxima $T_s$ and $T_m$ (un-substituted samples) and $x_c$ for several *Ln*FeAs(O$_{1-x}$F$_x$) systems.

Table 4.1: Structural $(T_s)$ and magnetic $(T_m)$ transformation temperatures with critical amount of F-substitution suppressing the structural transformation $(x_c)$ observed in various *Ln*FeAs(O$_{1-x}$F$_x$) systems.

| *Ln* | $T_s$ (K) | $T_m$ (K) | $x_c$ | Ref. |
|------|-----------|-----------|-------|------|
| La | $150 - 160$ | 132-138 | $0.045 - 0.08$ | [172,191,192,193,194,195,196,197] |
| Ce | $145 - 158$ | | 0.06 | [198,199] |
| Pr | $153 - 154$ | $127 - 130$ | 0.08 | [200,201] |
| Nd | 130 - 150 | $140 - 141$ | 0.10 | [202,203,204,205,206] |
| Sm | 130 - 175 | 133 | $0.045 - 0.14$ | [172,207,208,209] |
| Gd | 135 | 128 | | [172] |
| Tb | 126 | 122 | | [172] |

The $T_s$ and $x_c$ data listed in Table 2.4.I are scattered in several cases, even for the same systems; these differences can in part be ascribed to the feeble orthorhombic distortion: in pure compounds, the *a*- and *b*-axes differ by only $\sim 1\%$, and this variance progressively decreases with the increase in F content. As a rule, the structural transformation is evaluated by the selective peak splitting of the $\{hh0\}$ reflections, but in the case of a slight structural distortion, the measured profile function can be greatly affected by the contribution of the instrumental profile function, preventing in some cases a resolved peak splitting and consequently the precise determination of $T_s$. An additional complication arises for electron-doped materials: samples prepared in different laboratories are often characterized by a different real F-content, despite having the same nominal composition. Therefore, remarkable variations of both $x_c$ and $T_s$ are reported for samples prepared and analyzed under different experimental conditions. Conversely, a general good agreement characterizes the values of $T_m$ for pure end-member compositions.

As a rule, the phase diagrams of $Ln$FeAs(O$_{1-x}$F$_x$) systems display some common features:

1) The $P4/nmm \rightarrow Cmme$ structural transformation is $1^{st}$ order for undoped $Ln$FeAsO. Unfortunately, no studies have been carried out in order to ascertain if and how the nature of the phase transition changes with increasing F-content.

2) The diagram displays a plateau in $T_s$ below $x_c$ and an abrupt fall of $T_s$ as $x_c$ is approached, down to the complete suppression of the structural transformation. Such behaviour complies with the occurrence of local strain fields: a local strain field is created in a solid solution by replacing an ion by a larger/smaller one, since the matrix locally deforms to accommodate the change. When a few isolated strain fields are present, they have no or only minor effects on $T_s$, but the bulk properties change as strains increase in number and begin to overlap [210]. The ionic radii of O$^{2-}$ and F$^-$ actually differ by only ~ 5% [211], even though the effect related to electron doping cannot be disregarded. From a chemical point of view, the O$^{2-}$ - F$^-$ substitution raises another question; in fact, charge compensation should be preserved after chemical substitution. It is possible that very few percents of Fe-vacancies could gain the charge balance of the sample. Evidence for such a phenomenon has not yet been shown for $Ln$FeAs(O$_{1-x}$F$_x$) systems, but many experiments demonstrated the occurrence of a strong amount of Fe-vacancies in the related K$_y$Fe$_{2-x}$Se$_2$ system [212,213,214].

3) The magnetic transition is well separated from the structural transformation. Static magnetism occurs only within the orthorhombic phase; this phenomenon suggests a close dependence of the magnetic state on lattice symmetry. In this context, it was proposed that magnetic frustration takes place and is partially relieved when the tetragonal symmetry is broken, whereas another scenario foresaw that both transitions are driven by orbital ordering [9].

4) One of the most fascinating topics exhibited by Fe-based superconductors is the aforementioned relationship (coexistence or segregation) between the magnetic and superconductive phases. Early experiments in under-doped $Ln$FeAs(O$_{1-x}$F$_x$) compounds suggested a dependence on $Ln$ of the magnetic to superconductive ground state transition. In particular, a crossover from a $1^{st}$ order- like, to a quantum critical point, up to a $2^{nd}$-order-like transition was suggested, crossing over La-, Ce- and Sm-based systems, respectively [170]. The amount of results following earlier studies subverted this simple scenario, as contrasting evidence was reported by different research groups. A local electronic order in the Fe layer was suggested in under-doped systems, where low- and high-doped regions coexist, favouring superconductivity over static magnetism [215]. Separation was reported in the LaFeAs(O$_{1-x}$F$_x$) [192,195], even though nano-scale electronic inhomogeneities were later identified [215]. Separation was also reported in earlier studies for the CeFeAs($O_{1-x}F_x$) system [198], suggesting a possible quantum critical point [170], but later studies found a nanoscopic coexistence [216]. A slight overlap between the two electronic phase fields occurs in the SmFeAs($O_{1-x}F_x$) system due to nanoscopic coexistence [215,217,218,219,220], even though other studies argue a separation [208]. In any case, it seems quite clear that superconductivity and static magnetism can coexist within a limited under-doped compositional range, even though the nature of such coexistence deserves further analysis.

5) Superconductivity can initially arise in the under-doped orthorhombic phase, as shown in the LaFeAs($O_{1-x}F_x$) [195], CeFeAs($O_{1-x}F_x$) [198] and SmFeAs($O_{1-x}F_x$) systems [207]. We note that $T_c$ increases homogeneously with electron doping throughout the orthorhombic-to-tetragonal transformation, that is, $T_c$ exhibits no discontinuity at the crossover of the structural phase boundary.

### 4.1.1 The LaFeAs($O_{1-x}F_x$) system

In the beginning, the low-symmetry phase was said referred to crystallize in the monoclinic $P112/n$ space group ($P2/m$ space group in the standard setting) [191], but the correct *Cmme* structural model was introduced soon after [194]. A slight but evident decrease in $T_s$ with the increase in F-content up to $x \sim 0.04 - 0.05$ is reported by Luetkens *et al.* [192] and Huang *et al.* [195], whereas Qureshi *et al.* [196] suggest an almost constant value of $T_s$ up to $x = 0.045$. In the phase diagram drawn by Luetkens *et al.* [192], the orthorhombic structure and the static magnetism are both abruptly suppressed at the phase boundary of the superconducting state, even though a weak diffraction line broadening was observed during cooling for an under-doped superconductive composition, suggesting a wider amplitude of the orthorhombic phase field. Conversely, direct evidence for an orthorhombic distortion in the under-doped superconducting phase ($x = 0.05$) is reported in the phase diagram plotted by Huang *et al.* [195], implying that the evolution from the magnetic to superconductive ground states is not directly associated with the structural transformation.

The antiferromagnetic ordering of the Fe spins is characterized by the propagation wave-vector $\mathbf{k} = (1,0,\frac{1}{2})$, with anti-parallel nearest-neighbour spins along the $c$-axis, and an ordered moment of $\sim 0.3$ $\mu_B$ [191,197]; the magnetic moment results aligned along the $a$-axis ($b < a < c$ ) [196].

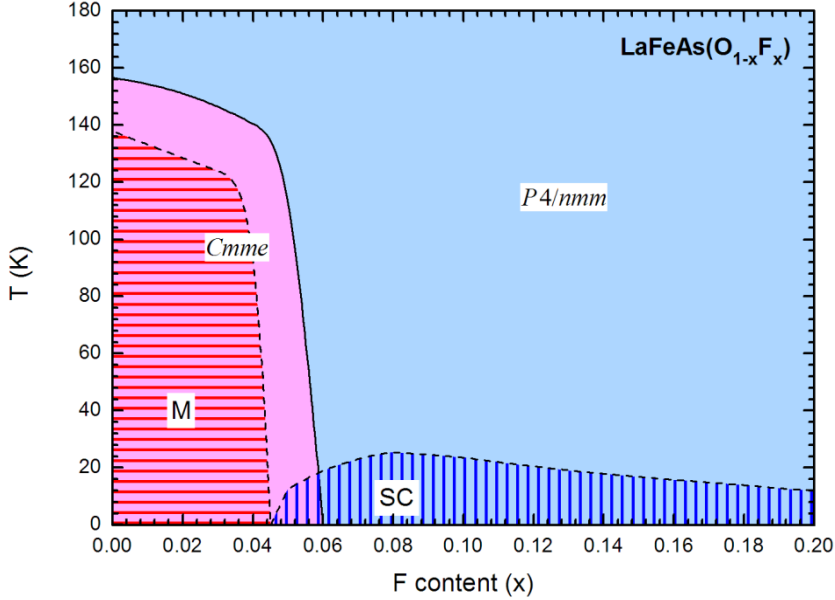

The LaFeAs(O$_{1-x}$F$_x$) system exhibits some peculiarities that are probably related to the size of the La$^{3+}$ ionic radius, which is the largest among $Ln^{3+}$ ions. In fact, the superconductive phase field exhibits a dome, but in the over-doped region, $T_c$ is progressively suppressed. Moreover, the superconductive state sets in as magnetic ordering is completely hindered, hence the two phases do not coexist under normal pressure, which is consistent with a 1$^{st}$ order quantum phase transition. On the other hand, hydrostatic pressure experiments demonstrated that these states can coexist under pressure, since they most probably are spatially separated in the crossover region of the phase diagram and competing for phase volume [221]; a similar mesoscopic segregation was obtained by chemical pressure in (La$_{0.7}$Y$_{0.3}$)FeAs(O$_{1-x}$F$_x$) samples [222].

### 4.1.2 The CeFeAs(O$_{1-x}$F$_x$) system

In the phase diagram drawn by Zhao *et al.* [198], $T_s$ progressively decreases with doping and static magnetism is suppressed before the arising of superconductivity; superconductivity first appears in the under-doped orthorhombic phase, with a relatively high onset $T_c$ (∼ 29 K).

After ordering, the magnetic moments of the Fe ions exhibit the same in-plane stripe structure as in LaFeAsO, but a parallel coupling of nearest-neighbour spins along the *c*-axis [198], which is consistent with the propagation vector **k** = (1,0,0) [9]. The ordered moment at the Fe sub-lattice is ∼ 0.8 $\mu_B$, whereas Ce magnetic moments order antiferromagnetically below ∼ 4 K [198].

Interestingly, both the magnetic and the superconducting order parameters are suppressed at the magnetic - superconductive boundary, suggesting the possible presence of a quantum critical point. This scenario is however questioned by the later results of Sanna *et al.* [216] and Shiroka *et al.* [223], which ascertained a nanoscopic coexistence between short-range magnetism and superconductivity for under-doped compositions. A significant transition from long- to short-range static magnetism was also detected with increasing doping, accompanied with or induced by a drastic reduction of the magnetic moment of the Fe ions.

### 4.1.3 The PrFeAs($O_{1-x}F_x$) system

Kimber *et al.* [224] report rather low values for both $T_s$ and $T_m$ (136 K and 85 K, respectively), notably different from those measured by Zhao *et al.* [201] as well as Rotundu *et al.* [200] (153 - 154 K and 127 - 130 K, respectively). In the phase diagram drawn by Rotundu *et al.* [200], superconductivity and magnetism do not coexist and the under-doped superconductive composition is tetragonal, even though a marked diffraction line broadening develops during cooling, accompanied by a net decrease of the peak intensity [200].

The spin ordering characterizing the Fe sub-lattice is the same as in CeFeAsO, with a magnetic propagation vector $\mathbf{k}$ = (1,0,0); Fe spins ($\sim$ 0.5 $\mu_B$) order antiferromagnetically along the *a*-axis and ferromagnetically along the *b*- and *c*-axis, with the magnetic moment aligned along the *a*-axis ($b <$ $a < c$ ) [201,224]. Below 12-14 K, the magnetic moments at the Pr sub-lattice order antiferromagnetically, as in CeFeAsO [201,224].

### 4.1.4 The NdFeAs($O_{1-x}F_x$) system

This system is rather unexplored at present: only one phase diagram is available and the interplay between the electronic phases has not yet been studied in detail. The phase diagram plotted by Malavasi *et al.* [204] shows that $T_s$ remains substantially constant up to relatively high values of F-content ($x \sim 0.11$); the structural transformation is then abruptly suppressed by "a very few increase of substitution ($x \sim 0.13$)" and at the same time superconductivity arises.

Magnetic ordering of the Fe atoms in NdFeAsO [205] is the same as the one observed in LaFeAsO; it is characterized by an antiferromagnetic structure whose propagation wave-vector is $\mathbf{k}$ = (1,0,½) and for which the ordered moment is $\sim$ 0.25 $\mu_B$.

### 4.1.5 The SmFeAs($O_{1-x}F_x$) system

In the phase diagram drawn by Margadonna *et al.* [207], the superconductive state emerges in a rather wide, under-doped orthorhombic phase field. $T_s$ increases homogenously, crossing over from the orthorhombic to the tetragonal structure. In roughly the same wide under-doped orthorhombic superconductive regime, the electronic phase diagram plotted by Drew *et al.* [218] shows the clear coexistence of static magnetism and superconductivity. A later electronic phase diagram, reported by Sanna *et al.* [220], displays a rather abrupt crossover between the two electronic phases, which are confined within an extremely narrow doping range; such a different behaviour was ascribed to the better chemical homogeneity of the analyzed samples. In any case, the nanoscopic coexistence of magnetism and superconductivity has been ascertained by several independent works [218,219,220]. In the phase diagram constructed by Kamihara *et al.* [208], magnetism and superconductivity share a very limited compositional range, even though the authors state that such coexistence is actually only apparent, and must be ascribed to crystallographic and/or compositional disorder. In this diagram, superconductivity also emerges in the under-doped orthorhombic phase as found by Margadonna *et al.* [207], but the homogeneity range of the orthorhombic phase is much more limited. A quantum critical point arising from the competition between the antiferromagnetic and superconductive ground states has been debated to occur at $x \sim 0.14$ [217].

In SmFeAsO, the $Sm^{3+}$ moments order antiferromagnetically below $\sim 5$ K, and with F-substitution, the transition decreases slightly down to $\sim 4$ K in the superconductive SmFeAs(O$_{0.85}$F$_{0.15}$) [225]; this magnetic transition thus appears to be almost insensitive to the orthorhombic-to-tetragonal structural transformation that occurs with doping. In the orthorhombic and tetragonal phases, the $Sm^{3+}$ spin ordering corresponds to the *Cm'm'e'* and *P4/n'm'm'* Shubnikov space groups, respectively [226].

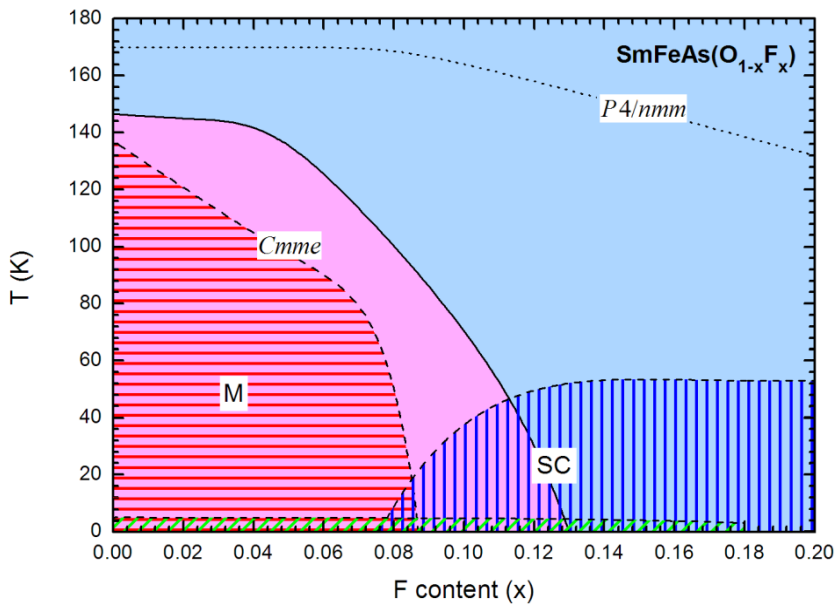

Figure 4.2: Assessed phase diagram of the SmFeAs(O$_{1-x}$F$_x$) system (reference data from ref. [189,207,208,219,220]; the hatched region below 5 K represents the phase field where the $Sm^{3+}$ spins order. The dotted line represents the equilibrium curve for the structural transformation estimated by analyzing the thermal dependence of the selective line broadening [227].

A rather different structural phase diagram was proposed by Martinelli *et al.* [227] (Figure 4.2), where even at optimal doping, F-substitution only suppresses static magnetism, but not the orthorhombic distortion. Such a different result was achieved by assuming that the contribution of instrumental resolution prevents a resolved splitting of the observed peak for very reduced structural distortions; hence the selective diffraction line broadening (coupled with an abrupt intensity decrease) developing in under- and optimally-doped samples during cooling was treated as an effect originating from a highly reduced orthorhombic lattice distortion. Remarkably, the retention of the structural transition in optimally doped samples was subsequently confirmed by an independent [19]F NMR investigation [228]. In this context, it is worth noting that similar selective line broadening/line intensity decreases can be detected by careful inspection of many diffraction data published for other (tetragonal alleged) doped 122- and 1111-type compounds.

## 4.2 Electron-doped systems: H-substitution

The $Ln$FeAsO, as well as the $AE$FeAsF compounds, can be electron-doped by H-substitution; hydrogen occurs as $H^-$ and as a consequence, superconductivity can be obtained in $Ln$FeAsO, but not in $AE$FeAsF compounds [229]. In particular, the LaFeAs($O_{1-x}H_x$) system is characterized by the presence of two distinct domes in the superconductive phase field. The first one is roughly coincident with that observed in the LaFeAs($O_{1-x}F_x$) system, whereas the second dome is characterized by a higher $T_c$ [230]; under pressure ($p = 3$ GPa), these two domes merge into a unique wider dome with maximum $T_c = 47$ K [230]. Conversely, in the phase diagrams of the SmFeAs($O_{1-x}H_x$) and CeFeAs($O_{1-x}H_x$) systems, a single dome is observed. We note that the magnetic and superconductive phase fields overlap perfectly in F- and H- substituted SmFeAsO and CeFeAsO systems [229,231]

## 4.3 Hole-doped ($Ln_{1-x}AE_x$)FeAsO and LnFeAs($O_{1-x}\square_x$) systems

Hole doping in ($Ln_{1-x}AE_x$)FeAsO compounds is obtained by partially replacing $Ln^{3+}$ with $AE^{2+}$ ions or by introducing vacancies at the O sub-lattice [232], whereas for $AE$FeAsF compositions, it occurs by partial substitution of F with O. No phase diagrams are available for these systems.

In any case, structural analyses carried out on ($Nd_{1-x}Sr_x$)FeAsO compounds showed that the structural transformation is also retained at optimal doping, with $T_c$ remaining almost constant with the increase in Sr-content [233]; this result strongly resembles that obtained by Martinelli *et al.* [227] for the SmFeAs($O_{1-x}F_x$) system. The highest $T_c$ for ($Ln_{1-x}AE_x$)FeAsO compounds generally ranges around ~ 15 K [233,234,235,236], except for the case of ($La_{1-x}Sr_x$)FeAsO, where a $T_c$ as high as ~ 25 K is measured at optimal doping [237].

The preparation of hole-doped $Ln$FeAs($O_{1-x}\square_x$) compounds requires high-pressure synthesis techniques. The superconductive properties characterizing these systems are comparable with those measured in the homologous F-substituted ones. The $T_c$ increases with the increase of the atomic number of $Ln$ from La to Nd, then stabilizes around ~ 53 K for $Ln$ = Nd, Sm, Gd, Tb, Dy [238,239]. A systematic study of the NdFeAs($O_{1-x}\square_x$) system showed that the maximum $T_c$ in its dome-shaped superconductive phase field is attained at an equivalent doping level in both NdFeAs(O$_{1-x}$F$_x$) and NdFeAs(O$_{1-x}$F$_x$) systems [240].

## 4.4 Transition metal substitution

### 4.4.1 Mn substitution

In principle, Mn substitution should act as hole-doping, but it is actually detrimental to superconductivity; such an effect becomes astonishing in La-based systems. In fact, extremely low amounts of Mn (as low as $x = 0.002$) lead to the complete suppression of the superconductive state, whereas static magnetism sets in for $x > 0.001$ in La(Fe$_{1-x}$Mn$_x$)As(O$_{0.89}$F$_{0.11}$) samples [241]. It has also been debated whether a quantum critical point is present at the boundary between the superconductive and magnetic ground states [241]. In Nd- and Sm-based systems, superconductivity is also suppressed, but the amount of Mn-substitution is more than 10 times larger [242,243].

### 4.4.2 Co-substitution

Electron doping obtained by Co-substitution introduces the carriers directly in the FeAs layer, but is also detrimental to superconductivity, since it produces disorder at the Fe-plane, hence the highest $T_c$ never exceeds 20 K in $Ln$(Fe$_{1-x}$Co$_x$)AsO systems. This disorder produces a decrease in $T_c$ in the over-doped regime; the phase diagrams of these systems thus display a dome-shaped superconductive phase field, as reported for the La(Fe$_{1-x}$Co$_x$)AsO [244], Ce(Fe$_{1-x}$Co$_x$)AsO [245,246], Pr(Fe$_{1-x}$Co$_x$)AsO [247], Nd(Fe$_{1-x}$Co$_x$)AsO [251,248], Sm(Fe$_{1-x}$Co$_x$)AsO [244,249,250] and Gd(Fe$_{1-x}$Co$_x$)AsO [246] systems. This is the only apparent feature distinguishing the $Ln$(Fe$_{1-x}$Co$_x$)AsO systems from the $Ln$FeAs(O$_{1-x}$F$_x$) ones. The tetragonal-to-orthorhombic structural transformation decreases in temperature with the increase of Co-content, and is completely suppressed in optimally substituted samples [251,252], even though superconductivity first arises in the under-doped orthorhombic phase in both Nd(Fe$_{1-x}$Co$_x$)AsO and Ce(Fe$_{1-x}$Co$_x$)AsO systems [251,253]. With regard to the relationship between the magnetic and superconductive phases, in the La(Fe$_{1-x}$Co$_x$)AsO system, superconductivity emerges at $x \sim 0.025$ where magnetism is already completely suppressed, and no phase coexistence between magnetism and superconductivity seems to be taking place [244], as in the homologous LaFeAs(O$_{1-x}$F$_x$) system. Instead, conflicting results are reported for both Sm(Fe$_{1-x}$Co$_x$)AsO [244, 249] and Ce(Fe$_{1-x}$Co$_x$)AsO systems [246]. In-depth analyses actually ascertained equivalent coexistence of magnetism and superconductivity in both Sm(Fe$_{1-x}$Co$_x$)AsO [254] and Ce(Fe$_{1-x}$Co$_x$)AsO [245] systems within a narrow compositional range. In particular, a crossover occurs in the Ce(Fe$_{1-x}$Co$_x$)AsO system, from a long- to a short-range magnetic order, where the superconductive phase segregates [245]; this behaviour is fully consistent with that observed in the homologous $CeFeAs(O_{1-x}F_x)$ system [223]. A similar mesoscopic separation between magnetism and superconductivity phases has also been observed for under-doped $Ca(Fe_{1-x}Co_x)AsF$ compositions [173,255,256].

### 4.4.3 Ru-substituted systems

Ruthenium is iso-electronic with iron, but in the 1111-type compounds, Ru atoms sustain no magnetic moment, and the Ru-Fe substitution progressively frustrates magnetism [257]. In the 122-type compounds, superconductivity can be achieved by diluting the Fe layer with non-magnetic Ru, but not in in $Ln(Fe_{1-x}Ru_x)AsO$ systems [179,180,258]. In the $La(Fe_{1-x}Ru_x)AsO$ system Ru-substitution progressively suppresses the structural transformation and the magnetic transition; a crossover from a $1^{st}$ to a $2^{nd}$ order character of the structural transformation takes place with substitution [179], and the symmetry breaking is completely suppressed for $x \gtrsim 0.4$ [179,180], whereas magnetism at the Fe sub-lattice is destroyed for $x \gtrsim 0.6$ [179,180,258]. With the suppression of the orthorhombic phase, antiferromagnetic ordering in the tetragonal phase becomes short-ranged; such unusual magnetic state seems to be related to the occurrence of a lattice strain in the tetragonal lattice [179]. We note that in optimally electron doped $Ln(Fe_{1-x}Ru_x)As(O_{1-x}F_x)$ systems, Ru-substitution induces a re-entrant static magnetic phase that nanoscopically coexist with the superconductive phase. Moreover, with the onset of static magnetism, a marked decrease of $T_c$ occurs, indicating competition between the two order parameters [259,260].

### 4.5 P-substitution

Superconductivity in Fe-SC compounds was first discovered in LaFePO with $T_c$ as low as ~ 4 K [261]. In addition, P-substitution in the $LaFe(As_{1-x}P_x)O$ system produces chemical pressure and can induce superconductivity ($T_c$ ~ 10 K) for $x = 0.25$–$0.30$ [262].

The $CeFe(As_{1-x}P_x)O$ system is outstanding among the 1111-type Fe-SC, due to the peculiar interplay between the $Fe^{2+}$ and $Ce^{3+}$ magnetic lattices. In fact, an antiferromagnetic to ferromagnetic ordering of $Ce^{3+}$ moments is observed at $x = 0.30$, as the magnetism at the Fe sub-lattice is weakened [176,178,263] and superconductivity ($T_c$ ~ 4 K) is reported to be close to this transition, possibly coexisting in a small homogeneity range ($x$ ~ 0.30) as a separated phase with static short-ranged antiferromagnetism at the Fe sub-lattice [178]. In this context, it is worth noting that initial studies found no evidence for superconductivity in the $CeFe(As_{1-x}P_x)O$ system [176,177]. The structural transformation and the antiferromagnetic ordering characterizing the CeFeAsO composition are suppressed by P-substitution around $x$ ~ 0.4, suggesting the presence of a magnetic quantum critical point [177], although this conclusion has been criticized after [178]. At a higher degree of substitution ($x \geq 0.95$), a heavy-fermion like behaviour takes place and the CeFePO end-member is a heavy-fermion compound [176].

In the SmFe(As$_{1-x}$P$_x$)O system, P-substitution suppresses magnetism [264], whereas the occurrence of superconductivity is still debated. At first, a superconductive state was reported to occur in a narrow homogeneity range with $0.5 < x < 0.65$ (maximum $T_c = 4.1$ K) [174], but later studies revealed that superconductivity can occur only if vacancies are present in the O sub-lattice [264].

## 5. Comparison with the phase diagrams of other unconventional superconductors

A comparison with the phase diagrams of other superconductors, in which the pairing mechanism is not mediated by phonons, is needed and would be instructive. As already stated, the phase diagrams of the Fe-SC and other unconventional superconductor systems, such as cuprates and heavy-fermion superconductors, are closely similar [265]. In particular, the $T$-$p$ phase diagrams of the heavy-fermion CePd$_2$Si$_2$ [266] and doped BaFe$_2$As$_2$ compounds look impressively similar (Figure 5.1). We note that the two-parent compounds are isostructural, suggesting that the underlying structural properties play a primary role. In all of these systems a magnetic ground state is present and only after its weakening or suppression superconductivity can develop. It is not yet known how similar the superconducting mechanism is in these systems, but the proximity between the superconductive and magnetic states appears to be a fundamental prerequisite for high-temperature superconductivity. We note that the crystal structure of the aforementioned heavy-fermion CePd$_2$Si$_2$ compound belongs to the tetragonal system, while the symmetry of its magnetic unit cell is orthorhombic (magnetic space group: $F_C m'm'm$) [267]. The magnetic nature of the symmetry-breaking anisotropy could indicate that spin nematic order is even present in the physics of some heavy-fermion compounds.

The presence of a spin-density wave state has been experimentally established in heavy-fermion compounds and several Cu-SC systems. Moreover, spin density wave and nematicity emerged in the last few years as the fundamental component of the physics of Cu-SC compounds in the pseudo-gap state [268,269]. In particular, recent investigations on the YBa$_2$Cu$_3$O$_{7-x}$ system ascertained that even in this system a competition between charge and spin density wave states with superconductivity takes place, with a crossover from the density wave state to the superconductive one strongly resembling the crossover observed in Fe-SC [269].

Nonetheless the Fe-SC and Cu-SC systems display some significant differences. The parent compound of Cu-SC is a Mott insulator where magnetism is driven by local moments, whereas the parent compound of Fe-SC is a semi-metal with a magnetic state that is probably related to itinerant electrons. In Cu-SC, superconductivity is induced by charge doping, suppressing magnetism; in particular, in Cu-SC materials, the appearance of superconductivity seems to be strictly connected with the valence of Cu, and $T_c$ reaches its maximum value when the valence of Cu reaches $\sim 2.2$ v.u. [270]. Conversely, in Fe-SC, superconductivity can also be obtained without changing the carrier concentration, for example by applying pressure, by isovalent substitution and, in some cases, by diluting the Fe sub-lattice with a non-magnetic species such as Ru. For Cu-SC materials, the electron- and hole- doped sides of the phase diagram exhibit significant differences. Indeed, a pseudo-gap region separating the magnetic and superconducting phase fields is found in the hole-doped side of the phase diagram, whereas its occurrence in the electron-doped side is debated. This pseudo-gap region is however not present in Fe-SC phase diagrams. Superconductivity in Fe-SC is quite resistant to chemical substitution at the Fe sub-lattice, whereas a very low concentration of substitutional dopants at the Cu sub-lattice induces the complete suppression of superconductivity in Cu-SC. Finally, the superconducting dome is rather symmetric in Cu-SC, but not so in Fe-SC materials, and the superconducting order parameter has a $d$-wave symmetry in Cu-SC.

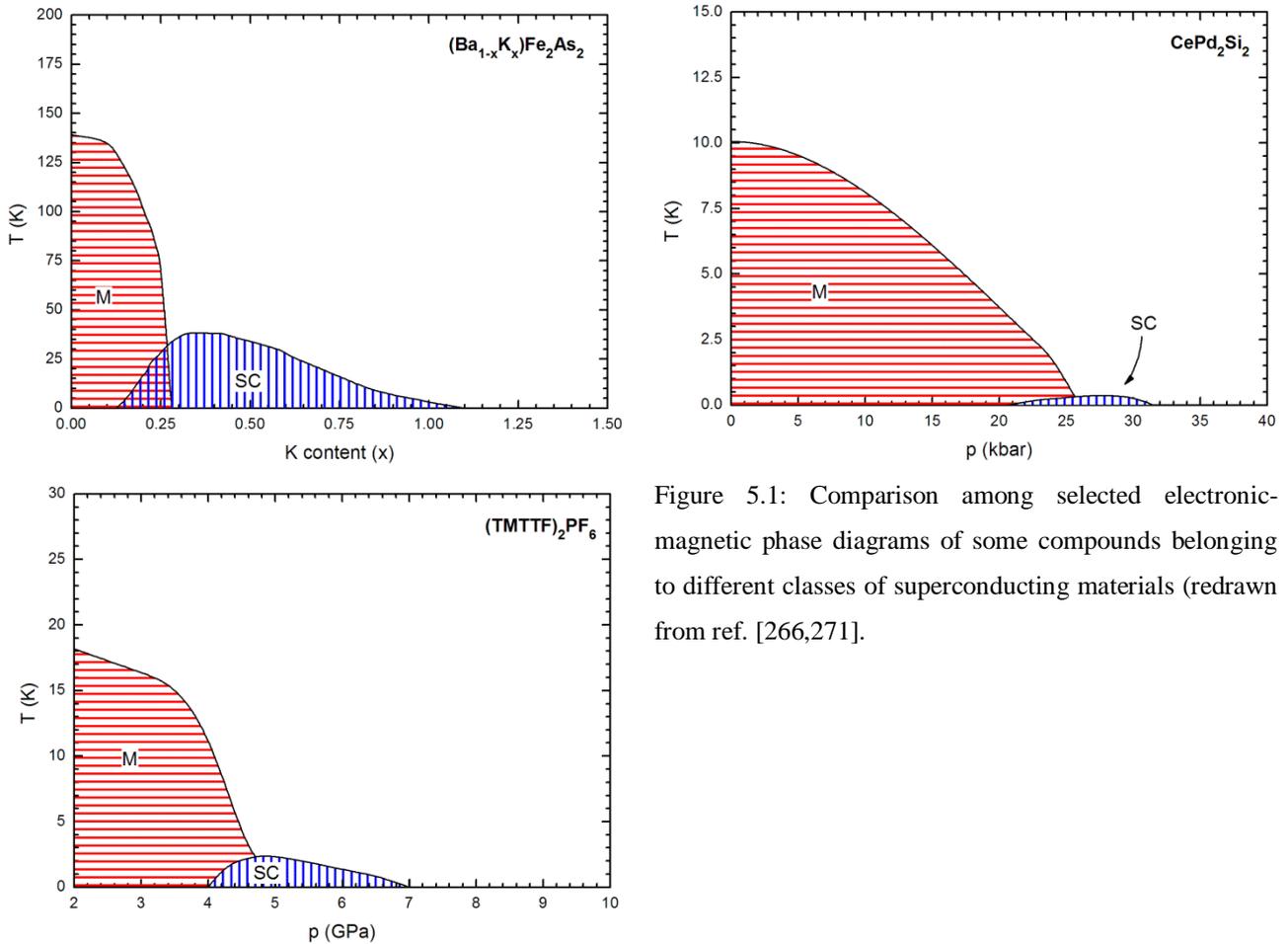

Figure 5.1: Comparison among selected electronic-magnetic phase diagrams of some compounds belonging to different classes of superconducting materials (redrawn from ref. [266,271].

Even the phase diagrams of some organic superconductors display similarities with those of Fe-SC. For example, a crossover from a spin density wave towards a superconducting ground state occurs under pressure in the $(TMTSF)_2PF_6$ Bechgaard salt (Figure 5.1). In particular, the inter-relationship between the spin density wave and superconducting phase field in the corresponding phase diagram is impressively similar to that characterizing the phase diagram of 122-type compounds [271].

Finally, it is worth remembering that at very high pressures, even the phase diagram of pure Fe displays a superconductive dome-shaped field, between ~15 and 30 GPa (highest $T_c$ ~ 2 K at ~ 21 GPa) [272]. In this case, the appearance of superconductivity is related to the cubic − hexagonal structural transformation taking place around ~ 10 GPa. In fact, the cubic structure is stable at lower pressures and is ferromagnetic, while in the high-pressure hexagonal phase, there is growing evidence for a weak antiferromagnetic state [273]. More interestingly, the pairing mechanism seems to be mediated by magnetic spin fluctuations, exactly like in Fe-SC materials.

## 6. Conclusions

In this review, we present and critically discuss the phase diagrams of the most extensively studied Fe-based 11-, 122- and 1111-type systems, seeking to provide a correlation between experimental evidence and theoretical models. Within the Fe-SC class of materials, chalcogenides and pnictides systems display rather different structural and magnetic properties. Notwithstanding, superconductivity seems to have a common origin that is induced by $(\pi,\pi)$ spin fluctuations. In principle, the structural transformation that is breaking the tetragonal symmetry can originate from lattice (phonons) or electronic (spin, charge or orbital order) degrees of freedom. A closer analysis of the phase diagrams, coupled with more specific experimental evidence, can give some clues for resolving this issue.

Within the Fe-SC class of compounds, the mechanism of the structural transformation characterizing $\beta$-$Fe_{1+y}Te$ on cooling is peculiar, since lattice and spin degrees of freedom interact cooperatively, giving rise to a coupled magneto-structural transition. Conversely, the lattice degrees of freedom do not play a major role in the transformation that is breaking the tetragonal symmetry in $\beta$-$FeSe_{1-x}$, 122- and 1111-type compounds [7]. In this context, the case of $\beta$-$FeSe_{1-x}$ also seems peculiar, since the structural transformation is not followed by magnetic ordering. The multitude of 122- and 1111-type compounds displaying a strict coupling between the structural and magnetic transitions biased a large number of theoretical studies towards a scenario where magnetism plays a primary, or even fundamental, role in the structural transformation. On this basis, it has been proposed that Ising-nematic fluctuations of magnetic origin drive the orthorhombic distortion. In this scenario, the long-range Ising-nematic order triggers orbital order and interacts cooperatively with it [4,274,275]. The proposed mechanism first involves the breaking of the $Z_2$ (Ising) symmetry in the temperature region between $T_s$ and $T_m$, where nematic fluctuations induce the breaking of the tetragonal symmetry. The breaking of the $O(3)$ spin-rotational symmetry then follows, inducing static antiferromagnetic ordering [274]. The nematic model can explain some features of the anisotropic behaviour of Fe-SC. For instance, calculations of the resistivity anisotropy, based on the experimentally observed splitting between the on-site orbital energies, give the opposite sign for the experimental resistivity anisotropy. This suggests that orbital ordering alone cannot explain the observed resistivity anisotropy. Moreover, nematic fluctuations are used to explain the softening of the shear modulus even at high temperatures. On the other hand the nematic order competes with SC, which can explain the re-entrance of the paramagnetic phase inside the superconducting dome of some Fe-SC, as well as the suppression of the orthorhombic distortion below $T_c$. The nematic fluctuation origin of the tetragonal symmetry breaking complies with most of the experimentally observed features, while it hardly conforms to $\beta$-FeSe$_{1-x}$ and several under-doped 122- and 1111-type compositions in the phase diagrams, which are characterized by an orthorhombic structure and a fully superconductive state in which static magnetism is completely suppressed.

An alternative model predicts that electron nematicity originates from orbital degrees of freedom: the occurrence of a ferro-orbital order induces the orthorhombic lattice distortion and triggers the magnetic-ordered state [1,274,275]. This model complies better with the phase relationships that are observed in the phase diagrams. Recent analyses on $\beta$-FeSe$_{1-x}$ support this scenario in which orbital degrees of freedom drive the structural transformation and compete with superconductivity [53,276,277]. Indeed, ARPES measurements on pure and doped BaFe$_2$As$_2$ are also consistent with the Fermi surface topology that is predicted in the orbital-ordered states [18,278,279,280]. Thus, experimental results indicate that electron nematicity is likely driven by orbital degrees of freedom, even though theoretical analyses suggest that the ferro-orbital and nematic orders are cooperative instabilities, leading to the enhancement of both $T_s$ and the anisotropic properties of the orthorhombic state [274]. Within this scenario, the antiferromagnetic ordering characterizing Fe-pnictides is triggered by the orthorhombic distortion of the lattice, and not *vice versa*. A few exceptions are found, for which magnetism takes place inside a tetragonal structure; a closer analysis reveals that in these cases, concurrent lattice microstrains are generally present or likely to occur, locally breaking the tetragonal symmetry [179].

It must be noted that some confusion persists in the literature about the characterization of the nematic phase. By definition, this phase entails a reduction of the rotational symmetry ($C_4 \rightarrow C_2$) that is preserving the translational symmetry; this corresponds to a tetragonal-to-orthorhombic structural transformation from the crystallographic point of view (in this light, the structural phase fields in almost all phase diagrams should be redrawn). Some authors emphasize the electronic origin of the structural transformation and adopt the term "nematic", even when the translational symmetry is broken and the lattice displays a net orthorhombic distortion.

On the other hand, the features of the experimental phase diagrams give no clues about the nature of the ordered magnetic state, that is whether it is related to a localized or itinerant character of the electrons.

A rather common feature exhibited by most phase diagrams is the microscopic coexistence between the antiferromagnetic and superconductive states within more or less restricted compositional ranges: the coexistence field between the antiferromagnetic and superconductive phases is rather wide in the 122-type compounds; in the 1111-type systems, it is quite narrow or even null, but in some cases, it can be induced by applying pressure. This microscopic phase coexistence is probably an intrinsic property of the 122- and 1111-type systems and suggests the existence of the $s^{+-}$ superconducting order. Under no circumstances magnetism and superconductivity are separated by a paramagnetic phase field, suggesting that both ground states compete for the same conduction electrons. Instead, the proximity to a magnetic quantum critical point suggests that the electron pairing could be produced by the same magnetic interactions driving the magnetic ordering.

A large amount of disorder can be accommodated at the Fe sub-lattice, but in all cases, a detrimental effect on superconductivity is always present. Indeed, the comparison of the $(Ba_{1-x}K_x)Fe_2As_2$ and $Ba(Fe_{1-x}Co_x)_2As_2$ systems, as well as the analysis of the $(Ba_{1-x}K_x)(Fe_{1-y}Co_y)_2As_2$ system clearly reveal that the maximum value of $T_c$ is reduced by the disorder induced by Co-substitution.

In conclusion, the characteristic features of the Fe-SC systems are rather well described by the schematic phase diagram drawn in Figure 1.1. Variations on this 'main theme' can be observed in specific phase diagrams, but it is reasonable to expect a universal phase diagram to hold for this class of materials. In addition, the relationship and interplay between the magnetic and superconductive ground states (one of the most fundamental and critical issues in Fe-SC) closely resembles the features characterizing other superconducting compounds belonging to very different classes of materials. This observation suggests that the suppression of long-range magnetic ordering, but not of the magnetic pairing in its entirety, can be a promising way to obtain unconventional superconductivity.

## Acknowledgments

A.M. acknowledges G. Lamura (SPIN-CNR) and I. Pallecchi (SPIN-CNR) for discussions and comments. A.M and F.B. acknowledge partial financial support from the FP7 European project

SUPER-IRON (Grant Agreement No. 283204); F.B. acknowledges partial financial support from PRIN2012 (no. 2012X3YFZ2_006).